\documentclass[10pt,journal,compsoc]{IEEEtran}

\usepackage{float}
\usepackage[style=base]{caption}
\usepackage{tabularx}
\usepackage{booktabs}
\usepackage[flushleft]{threeparttable}
\usepackage[pdftex]{graphicx}
\usepackage[table]{xcolor}
\usepackage[nocompress]{cite}
\usepackage{url}
\usepackage{amsmath}
\usepackage{siunitx}
\usepackage{jabbrv}

\DeclareGraphicsExtensions{.png,.pdf,.jpg}
\definecolor{Gainsboro}{cmyk}{0.00,0.00,0.00,0.14}

\newcolumntype{C}{>{\centering\arraybackslash}X}

\def\xtimes{\mkern-4mu\times\mkern-4mu}
\def\invivo{\emph{in vivo}}

\begin{document}

\title{MRSNet: Metabolite Quantification from Edited Magnetic Resonance Spectra With Convolutional Neural Networks}

\author{M.~Chandler,
        C.~Jenkins,
        S.\,M.~Shermer,
        and F.\,C.~Langbein%
\IEEEcompsocitemizethanks{\IEEEcompsocthanksitem M.~Chandler and F.\,C.~Langbein are with the School of Computer Science and Informatics, Cardiff University, Cardiff, UK.\protect\\%
E-mail: \text{hello@ma.ax}, frank@langbein.org%
\IEEEcompsocthanksitem C.~Jenkins is with CUBRIC at Cardiff University, Cardiff, UK.\protect\\%
E-mail: JenkinsC23@cardiff.ac.uk%
\IEEEcompsocthanksitem S.\,M.~Shermer is with the Department of Physics at Swansea University, Swansea, UK.\protect\\%
E-mail: lw1660@gmail.com}}

\IEEEtitleabstractindextext{
\begin{abstract}
Quantification of metabolites from magnetic resonance spectra (MRS) has many applications in medicine and psychology, but remains a challenging task despite considerable research efforts. For example, the neurotransmitter $\gamma$-aminobutyric acid (GABA), present in very low concentration \invivo{}, regulates inhibitory neurotransmission in the brain and is involved in several processes outside the brain. Reliable quantification is required to determine its role in various physiological and pathological conditions. We present a novel approach to quantification of metabolites from MRS with convolutional neural networks --- MRSNet. MRSNet is trained to perform the multi-class regression problem of identifying relative metabolite concentrations from given input spectra, focusing specifically on the quantification of GABA, which is particularly difficult to resolve. Typically it can only be detected at all using special editing acquisition sequences such as MEGA-PRESS. A large range of network structures, data representations and automatic processing methods are investigated. Results are benchmarked using experimental datasets from test objects of known composition and compared to state-of-the-art quantification methods: LCModel, jMRUI (AQUES, QUEST), TARQUIN, VeSPA and Gannet. The results show that the overall accuracy and precision of metabolite quantification is improved using convolutional neural networks.
\end{abstract}
}

\maketitle
\IEEEdisplaynontitleabstractindextext
\IEEEpeerreviewmaketitle

\IEEEraisesectionheading{\section{Introduction}\label{sec:introduction}}

\IEEEPARstart{M}{agnetic} resonance spectroscopy (MRS) is a rapidly developing, highly versatile tool, enabling us to acquire information about biochemical processes happening \invivo{} and to identify biochemical changes associated with various cancers and neurological disorders among others. The technique is non-invasive, painless, involves no ionizing radiation or radioactive tracers and can be performed on conventional clinical MRI scanners available in many hospitals. MRS has been responsible for advances in detecting, classifying and modelling malignant tissue in the brain~\cite{Garcia-Figueiras2016, Aggarwal2017, Zarinabad2018}, prostate~\cite{Kurhanewicz2008, Verma2010, Riches2015} and breast~\cite{Anna2009, Leong2015, Fardanesh2019}. It also has a role in modern psychology and neuroscience, where it provides a useful window into normal brain function, as well as abnormalities associated with conditions such as anxiety, schizophrenia, depression and other mood disorders~\cite{Strasser2019, DelliPizzi2016, Cordero2016, Taylor2017, Urrila2017}. Many of these studies have focused on the detection of $\gamma$-aminobutyric acid (GABA), the primary inhibitory neurotransmitter~\cite{McCormick2017}, as a key biomarker. Its quantification is especially difficult as it is present in low concentration \invivo{} and its characteristic MRS features overlap with those of much more abundant metabolites, such as N-Acetylaspartic acid (NAA) and Creatine (Cr), obscuring its signature in MR spectra. This had lead to the development of new spectroscopic techniques, such as edited spectroscopy, that attempt to selectively edit out certain features to make others observable. In this work, we focus on quantification of GABA with MEGA-PRESS, one of the most commonly used edited spectroscopy sequences.

An MR spectrum is a graph of the magnetic resonance signal $S(\nu)$ as a function of frequency $\nu$. MR spectra of different molecules have different features such as a set of characteristic peaks at distinct frequencies. While individual MR spectra are highly molecule-specific, identifying and quantifying the concentration of molecules from MR spectra of mixtures containing many molecules is a very challenging task as spectral features of different molecules overlap, signals from molecules present in low concentration may be barely distinguishable from the noise floor, resulting in low signal-to-noise ratios (SNR), and hardware or software calibration issues may distort the spectra obtained. In essence, quantification decomposes spectra into features associated with individual metabolites to infer the chemical composition of a sample or tissue voxel. There are many ways to tackle quantification, such as basis set fitting, singular value decomposition, principal component analysis or peak integration, focusing on either time or frequency domain analysis. In addition, spectral processing methods are employed for phase correction, removal of water or macro-molecule signals and baseline correction. Many of these require manual calibration and spectral processing may lead to distortion of spectra, thereby falling short of achieving accurate and precise quantification with minimal input from a human expert. \cite{Poullet2008}~is an excellent review of the state-of-the-art.

Basis set methods achieve decomposition by representing a spectrum as a linear combination of spectra from a basis set of individual metabolite signals and a residual signal. State-of-the-art methods generally employ a variant of least-squares fitting to match a linear combination to the spectrum. Creating basis sets itself, whether based on simulation or measurements, is not a trivial task. Further complicating basis set selection is that basis sets must usually be specifically created for the pulse sequence used to acquire the spectra, as the spectra depend on the pulse sequence. This is problematic as there are many pulse sequences and implementations on different systems can vary considerably~\cite{Chase2015}. Deriving basis sets from experimental data is also extremely time-consuming and the resulting basis sets may suffer from the same problems as the spectra to be quantified. Alternatively, using simulations for basis set generation relies on models, the accuracy of which has been questioned~\cite{Kreis2012}. Quantification, therefore, can be an extremely time-consuming task, where time must be spent acquiring or simulating basis sets as well as adjusting fitting and processing algorithms and parameters. This is where machine learning has an opportunity to improve quantification by providing a complete, unbiased solution in less time. Machine learning has already proved to be a useful tool in MR imaging~\cite{SelvikvagLundervold2018, Lee2018}, MR spectroscopy~\cite{Hatami2018, Lee2019a, Das2017, Hiltunen2002}, classification of Raman~\cite{Liu2017} and Faraday spectra~\cite{Brown2017}, and near-infrared (NIR) spectroscopy calibration~\cite{Cui2018a}.

We present MRSNet, a multi-class regression convolutional neural network (CNN) that can be trained quickly on a relatively small number of samples ($5,000$). We review MRS and associated quantification methods in Section~\ref{sec:background}. The data and method for the CNNs are discussed in Sections~\ref{sec:data} and~\ref{sec:method} respectively. We focus on frequency domain analysis by training the CNN on spectra to learn metabolite parameters, returning expected relative concentrations. Due to the time required to experimentally acquire a large number of spectra, the network is trained on a range of simulated data (see Section~\ref{sec:experiments}). As the results are heavily dependent on the accuracy of the basis set and the intra-basis set scaling, multiple basis sets from different sources are compared in Section~\ref{sec:data_source_comparison}. Furthermore, a range of network structures and different data representations of the frequency domain signals (real part, imaginary part and magnitude) are explored in Section~\ref{sec:data_type_and_channels}. The performance of MRSNet and other state-of-the-art methods is benchmarked using experimental spectra obtained from test objects (phantoms) of known composition (see Section~\ref{sec:evaluation}). The results show that on average CNNs are more accurate and precise. The experimental data used for the evaluation is available at~\cite{Data} and the code for the networks and spectral simulation is available at~\cite{Code}.

\section{Background}\label{sec:background}

We briefly review the acquisition of MR spectra and state-of-the-art quantification methods.

\subsection{Magnetic Resonance Spectroscopy}\label{sec:mrs background}

MRS utilizes the fact that elementary particles such as protons, neutrons and electrons have a quantum-mechanical degree of freedom called spin. As a result, atomic nuclei with an odd number of protons or neutrons such as hydrogen $^{1}$H have a net nuclear spin. $^{1}$H, the most abundant element in biological organisms~\cite{McQuarrie2009}, has two spin states. When an external magnetic field $B_0$ is applied, an energy gap of $h\nu$ develops between these states, where $h$ is the Planck constant and $\nu$ is the Larmor frequency, which is proportional to $B_0$. In a typical clinical MR scanner the Larmor frequency of $^{1}$H is in the radiofrequency (RF) range, e.g., \SI{127}{\mega\hertz} at \SI{3}{\tesla}. As the lower energy state is energetically preferred, a slight excess of nuclei are found in this state, leading to the development of a magnetic moment in the sample. This magnetic moment is almost undetectable, but when a suitable electromagnetic pulse at the Larmor frequency is applied, the spins can be rotated into a plane transverse to the external magnetic field. As the spins precess in this transverse plane they emit an MR signal while slowly returning to their equilibrium state, which can be detected by sensitive RF receive coils.

This time-domain signal is complex, as it has both magnitude and phase, and is Fourier-transformed to obtain the MR spectrum. The reason for this is that the features in the frequency-domain representation are directly related to the structure of the molecule the signal originates from. Although the Larmor frequency of an isolated proton depends only on the external $B_0$ field, protons in molecules have slightly shifted resonance frequencies due to chemical shielding effects, known as chemical shift. Chemical shifts can be reported in \si{\hertz} but it is customary to report chemical shifts in \si{ppm} (parts-per-million) with respect to the spectrometer frequency $\nu_0$. For a spectrometer calibrated to \SI{127}{\mega\hertz}, a chemical shift of \SI{1}{ppm} corresponds to \SI{127}{\hertz}. Chemical shifts accessible by MR spectroscopy are usually in the range of \SI{0}{ppm} to \SI{8}{ppm} and can be affected by pH and temperature of the sample. In addition to chemical shift, MR spectra are also influenced by $J$-couplings between nearby spins. This leads to splitting of the energy levels of the coupled spin system, which manifests itself as splitting of individual peaks into doublets, triplets, quadruplets, etc., depending on the number of spins that are simultaneously coupled. In a perfectly homogeneous field and in the absence of any relaxation the peaks would be $\delta$-function spikes. In practice, $T_1$ and $T_2$ relaxation of the precessing spins leads to line broadening and Lorentzian or Gaussian spectral lineshapes. This intrinsic line broadening is further increased by local variations in the $B_0$ field, resulting in a distribution of frequencies over an excited volume. A homogeneous $B_0$ field correlates with a narrower linewidth, sharper peaks and, importantly, reduced feature overlap between metabolite spectra.

In practice, MRS pulse sequences are more complicated than a single excitation pulse followed by immediate readout. Additional pulses must be applied to spatially localize the signal (Point RESolved Spectroscopy, PRESS)~\cite{BOTTOMLEY1987} and suppress unwanted signals such as the background water signal, e.g., by applying a CHEmical Shift-Selective (CHESS) pulse to saturate the water signal, followed by gradient dephasing~\cite{Haase1985}. Moreover, timing parameters such as $T_E$ (echo time; time between RF pulse and echo readout) and $T_R$ (repetition time; time between corresponding points in a repeating series of pulses and echoes) are adjusted to reduce residual magnetization between scans and macromolecular contributions, among other reasons. All of these choices affect the spectra obtained.

\begin{figure}
  \centering
  \includegraphics[width=0.48\columnwidth]{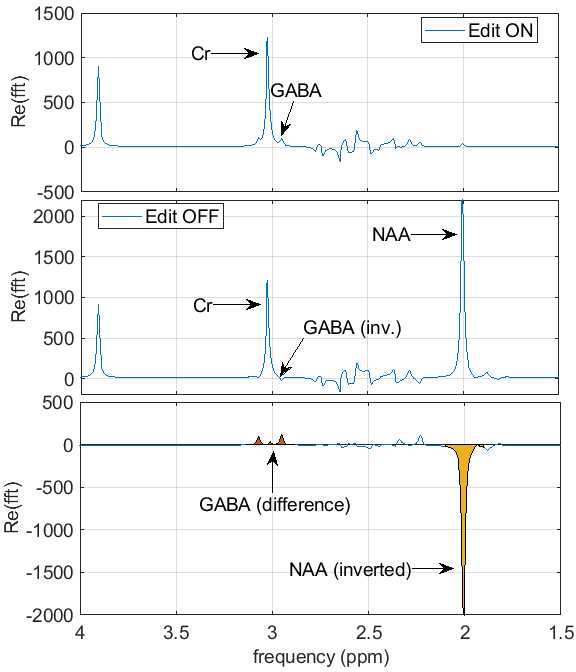}
  \includegraphics[width=0.48\columnwidth]{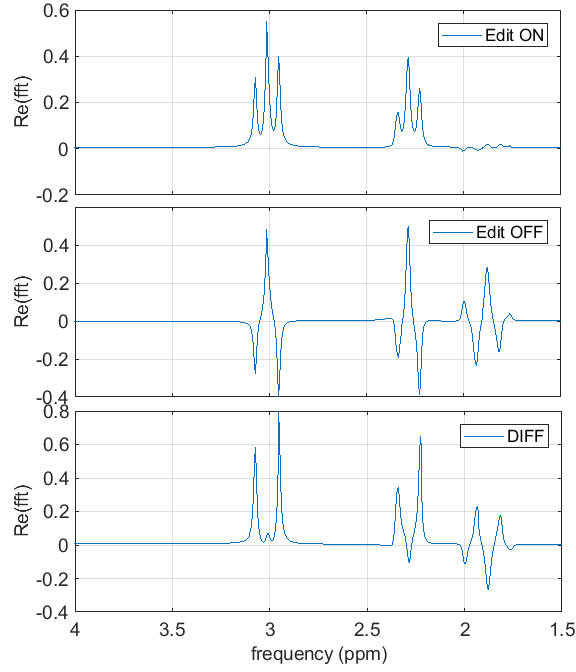}
  \caption{Simulated MEGA-PRESS spectra for mixture of \SI{15}{\milli\mole/\litre} NAA, \SI{8}{\milli\mole/\litre} Cr and \SI{3}{\milli\mole/\litre} GABA (left) and for GABA only, at much lower signal strength (right).}
  \label{fig:megapress_example_spectra}
\end{figure}

For edited spectroscopy, such as the MEGA-PRESS (MEsher-GArwood PRESS) sequence~\cite{Mescher1998}, additional editing pulses are applied between refocusing pulses to selectively enhance or suppress certain features. The standard MEGA-PRESS sequence produces two MR spectra, an edit-off and edit-on spectrum, from which a third difference spectrum is calculated. Specifically for GABA quantification, the editing pulses are applied at \SI{1.9}{ppm} during the edit-on acquisition and \SI{7.4}{ppm} during the edit-off acquisition. The aim of these editing pulses is the elimination of the Cr signal in the difference spectrum at \SI{3}{ppm}. As shown in Fig.~\ref{fig:megapress_example_spectra}, the small GABA signal at \SI{3}{ppm} is completely dominated by the much larger Cr signal. In the edit-off spectrum, the outer peaks of the GABA triplet at \SI{3}{ppm} are inverted compared to the edit-on spectrum, while the Cr signal is unaffected. In the difference spectrum, this ensures that the Cr signal at \SI{3}{ppm} is eliminated and only the side peaks of the \SI{3}{ppm} GABA triplet remain, amplified by a factor of $2$, making GABA detectable. The \SI{2}{ppm} NAA peak is absent from the edit-on spectrum as the \SI{1.9}{ppm} editing pulse also eliminates its corresponding magnetic moments.

\subsection{Quantification of MR Spectra}

Quantification of MR spectra aims to infer the relative contributions (concentrations) of different metabolites, accurately and precisely, even in the presence of environmental noise and low signal-to-noise ratios. While some metabolites are quite abundant and relatively easy to quantify, others of great importance, such as the inhibitory neurotransmitter GABA and the excitatory neurotransmitters Glutamine (Gln) and Glutamate (Glu)~\cite{Ramadan2013}, are considerably more challenging. This is due to low relative concentration \invivo{} and overlap of their spectral features with those of more abundant metabolites~\cite{Govindaraju2000}. For Glu/Gln the difficulty lies further in resolving their spectra, which are extremely similar at the field strength at which most clinical MRI scanners operate (\SI{1.5}{\tesla} or \SI{3}{\tesla}). The primary goal of this paper is the quantification of these difficult-to-quantify metabolites. As absolute quantification requires reliable reference spectra for calibration, which are generally not available for \invivo{} spectra, metabolite concentrations are typically reported as ratios, such as Cr/GABA, NAA/GABA and we also focus on relative rather than absolute quantification.

\begin{figure}
  \centering
  \includegraphics[width=\columnwidth]{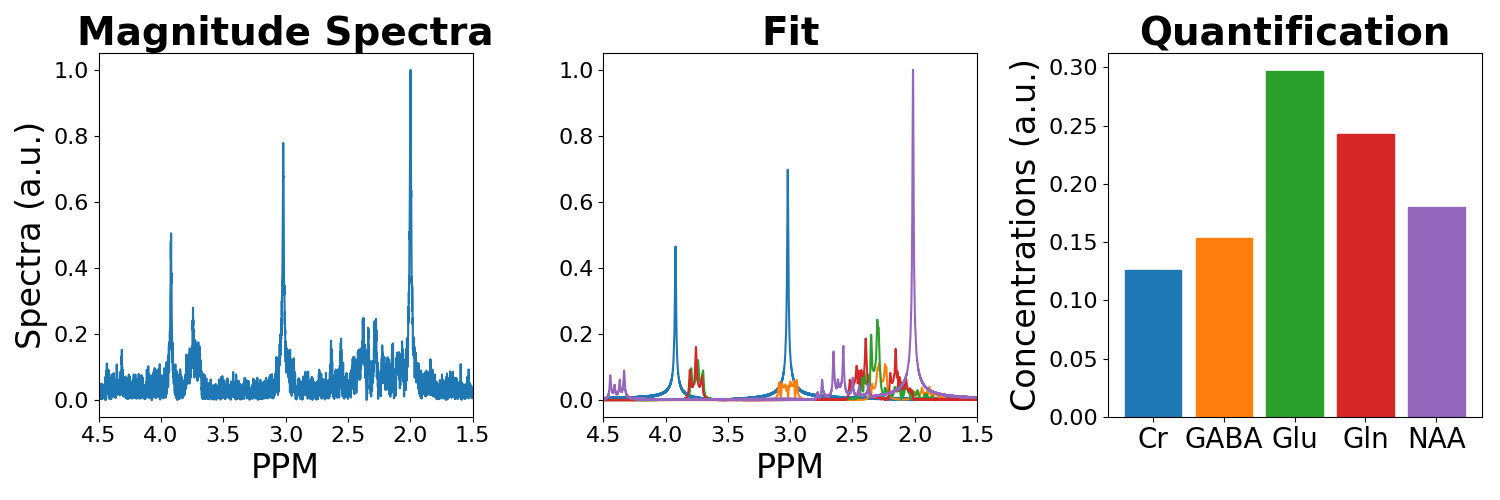}
  \caption{Simplified quantification example for spectrum containing Cr (blue), GABA (orange), Glu (green), Gln (red) and NAA (purple): simulated combined spectrum with noise (left), example fit (middle), quantification results (right).}
  \label{fig:quantification_example}
\end{figure}

\subsubsection{Traditional Approaches}

Various quantification methods have been developed, ranging from simple peak integration methods to fitting either the time or frequency domain signals with a linear combination of basis signals plus a residual in absolute or relative values as shown in Fig.~\ref{fig:quantification_example}. State-of-the-art methods typically focus on frequency domain analysis, due to difficulties interpolating time domain signals with varying acquisition parameters. The majority implement a variant of the non-linear least-squares (NLLS) algorithm to fit a set of parameters to a model. Methods that employ basis sets usually come bundled with a simulator to generate them. We briefly describe state-of-the-art software for quantification and simulation. For an excellent review see~\cite{Mierisova2001}.

LCModel~\cite{Provencher1993} is widely considered to be the gold standard of spectral quantification and uses NLLS to decompose spectra in the frequency domain. Basis sets are provided by the creator on request or can be obtained from suggested external sources. LCModel is only able to directly simulate macro-molecule contributions and not complete basis sets.

In contrast, TARQUIN~\cite{Wilson2011}, VeSPA~\cite{Soher2011} and jMRUI~\cite{Cabanas2009} are able to simulate their own basis sets for quantification. TARQUIN and VeSPA both perform NLLS fitting of frequency domain signals, while jMRUI has a range of algorithms available, including AMARES and AQSES for frequency domain analysis and QUEST for the time domain.

Gannet~\cite{Edden2011} focuses on peak identification and fitting specifically for edited MRS spectra, such as MEGA-PRESS, and does not use a basis set or simulator.

Despite the advances made in MR spectra quantification, it has been shown that there are significant errors in quantification between the different methods~\cite{Scott2016, Kanowski2004, Mierisova2001}. Recent attempts to compare and validate these tools for spectra obtained from carefully calibrated phantoms have shown very poor accuracy and large discrepancies between the concentrations reported by different tools~\cite{Jenkins2018, quantpaper}. These studies highlight the need for more reliable methods for quantification of MR spectra for MRS to become a reliable tool for clinical use, especially for difficult-to-quantify metabolites such as GABA.

Other methods utilising Bayesian inference have been developed for the analysis of time-domain signals in high-field NMR~\cite{Wilson2014}. These leverage prior knowledge via probability distributions and estimate `nuisance parameters' such as phase shifts, decay constants and noise parameters, outperforming Fourier based NMR methods. Due to much higher spectral resolution and the focus on molecular structure elucidation, the issues and applications of high-field NMR are different and the methods have not been applied to MRS quantification \invivo{}.

\subsubsection{Machine Learning Approaches}

Machine learning for MRS quantification has only recently begun to be explored. Das \emph{et al.}~\cite{Das2017} used random forests trained on $1$ million simulated spectra and tested on $287$ human subjects. More recently Lee \emph{et al.}~\cite{Lee2019a} trained CNNs on $40,000$ spectra to quantify a range of simulated and \invivo{} data. Both studies are focused on quantification with the single acquisition PRESS sequence, which generally precludes identification and quantification of difficult-to-detect metabolites such as GABA. Furthermore, the performance of the machine learning algorithms on \invivo{} spectra was assessed by comparing the quantification results with LCModel. Considering the large discrepancies in metabolite concentrations reported by different tools when provided with \emph{identical} input spectra, and the overall \emph{lack of accuracy} of the quantification results obtained with existing tools when tested against a wide range of spectra from calibrated phantoms, good agreement with LCModel results does not provide a strong validation of the method~\cite{Jenkins2018,quantpaper}.

Here we focus on the quantification of difficult-to-quantify metabolites in MEGA-PRESS spectra. The networks are trained on a comparatively small number of simulated spectra ($5,000$) and the results are validated independently using calibrated phantom spectra. A range of network structures, spectra representations, processing methods and the impact of the basis set choice are explored.

\section{Data}\label{sec:data}

Spectral datasets for quantifying mixtures of NAA, Gln, Glu, GABA and Cr are used due to their importance for MRS applications and the difficulty they pose for accurate quantification. These metabolites are in principle detectable by MEGA-PRESS. The networks are trained and validated using simulated data generated by combining simulated spectral basis sets into spectra based on the concentration of individual metabolites. Simulated spectral basis sets were deemed more efficient than deriving basis sets experimentally from calibrated phantoms. Using experimental data to generate basis sets is also problematic due to the difficulty of obtaining high-quality spectra~\cite{provencher2014lcmodel}. To assess the performance of the networks, experimental data from carefully calibrated phantoms are used as benchmark.

\subsection{Training and Validation Dataset Generation}\label{sec:simulation}

A basis set contains the normalised characteristic signals in the frequency domain for each metabolite, which are combined to create mixed spectra.

Basic simulators for basis sets often assume the molecule is prepared in a well-defined initial state such as the ground state. They approximate the pulse sequence by a series of instantaneous unitary operations acting on the state at certain times, followed by readout of the MR signal, to simulate the full MRS pulse sequence. More sophisticated simulators perform time-resolved calculations for finite-duration pulses and include the effects of relaxation or field inhomogeneities. However, the choices of pulse shapes and pulse timings in vendor-specific implementations of the MEGA-PRESS sequence differ~\cite{Chase2015} and are often not known precisely. Therefore, quite often educated guesses must be made. Similarly, other parameters, such as relaxation parameters or field inhomogeneities, are typically not known precisely. Moreover, even metabolite models, including chemical shift and $J$-coupling parameters, still have uncertainties~\cite{Kreis2012}. A major source of these parameters for common metabolites in MRS is the landmark paper by Govindraju \emph{et al.}~\cite{Govindaraju2000}, but alternative models have been suggested, e.g., for GABA by Near \emph{et al.}~\cite{Urban2010} and Kaiser \emph{et al.}~\cite{Song2008}.

Hence, we use and test different basis sets (see Section~\ref{sec:basis_comparison}) from state-of-the-art simulation and quantification software: FID-A~\cite{Simpson2017}, PyGamma~\cite{PyGamma} and LCModel~\cite{Provencher1993}. The MEGA-PRESS basis set for LCModel is generated by the MRS Lab at Purcell Health Sciences~\cite{Dr.JimMurdoch}. For PyGamma and FID-A, simulations are performed in house by adapting the MEGA-PRESS pulse sequence code provided. All simulations are aimed at the Siemens WIP MEGA-PRESS implementation, which was used to acquire the experimental datasets (see Section~\ref{sec:benchmark_datasets}). These basis sets were also used in a comparative study of state-of-the-art quantification methods~\cite{Jenkins2018, quantpaper} and the comparison with such methods in Section~\ref{sec:compare_with_current_methods}.

For each basis set, the training and validation datasets are generated by taking linear combinations of the basis elements with each metabolite having a scaling factor in $[0,1]$, corresponding to a relative concentration. The scaling factors are sampled using a low discrepancy (quasi-Monte Carlo) Sobol sequence~\cite{Sobol1967}, which provides good uniform coverage of possible states with a low number of data points. Time-domain noise is added from a normal distribution ($\mu=0$ and $\sigma$ randomly chosen in the range $[0,0.25]$) to $50\%$ of the dataset to improve simulation accuracy and network robustness. This noise model was chosen as it closely resembles what is seen in experimental spectra, by characterising the noise profile from spectral areas that do not contain a metabolic signal for $4,160$ phantom spectra.

\subsection{Experimental Benchmark Datasets}\label{sec:benchmark_datasets}

\begin{table*}
  \caption{Benchmark phantom composition: concentrations are in \si{mM} = \si{\milli\mole/\litre}.}
  \label{table:phantom_conc}
  \begin{tabularx}{\textwidth}{@{}lcc|ccccX@{}}
    Series & Medium & \# of Spectra & NAA    & Cr        & Glu    & Gln   & GABA \\
    \midrule
    E1     & Water  & $13$          & $15.0$ & $0$/$8.0$ &  $0.0$ & $0.0$ & $0.0$, $0.5$, $1.0$, $1.5$, $2.0$, $2.5$, $3.0$, $4.0$, $6.0$, $8.0$, $10.0$, $11.6$\\
    E3     & Water  & $15$          & $15.0$ & $8.0$     & $12.0$ & $3.0$ & $0.0$, $1.0$, $2.0$, $3.0$, $4.0$, $5.0$, $6.0$, $7.0$, $8.0$, $9.0$, $10.0$, $11.0$, $12.0$, $13.0$, $14.0$\\
    E4     & Gel    & $8$           & $15.0$ & $8.0$     & $12.0$ & $3.0$ & $0.0$, $1.5$, $3.0$, $5.0$, $7.0$, $10.0$\\
    \bottomrule
  \end{tabularx}
\end{table*}

\begin{figure}
  \centering
  \includegraphics[width=\columnwidth]{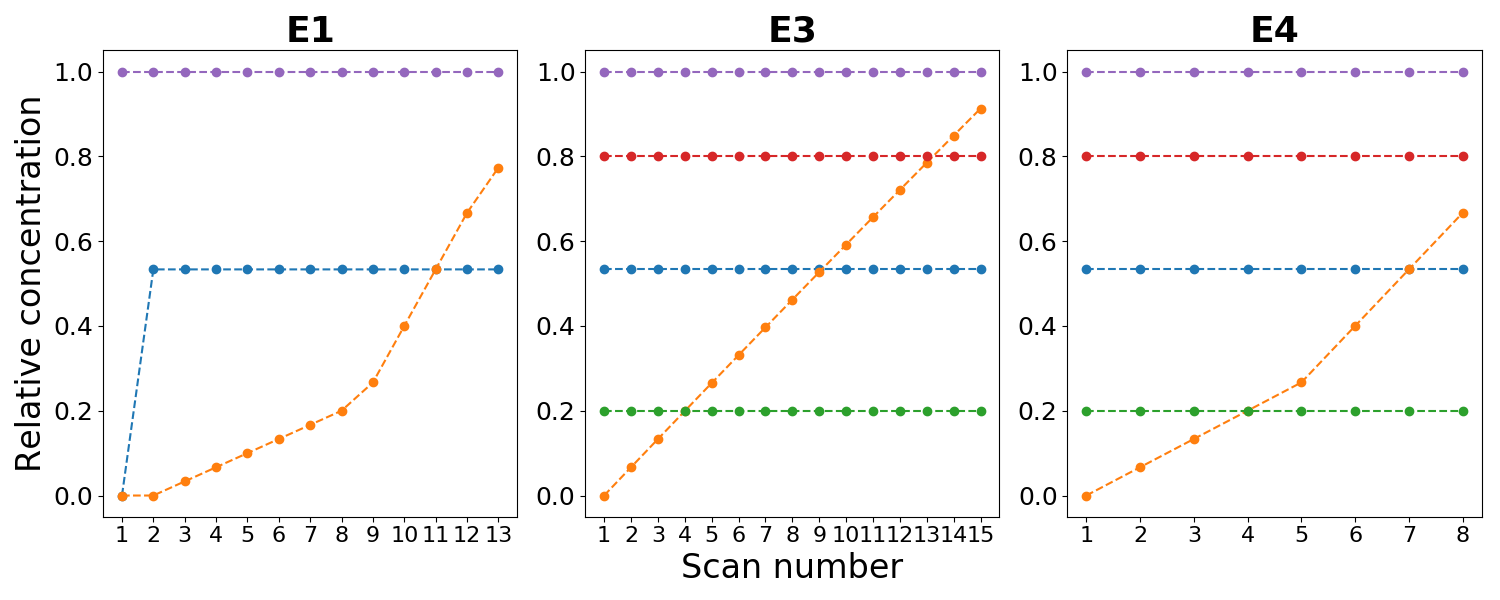}
  \caption{Metabolite concentrations relative to NAA, for datasets E1, E3 and E4 for NAA (purple: $1$), Cr (blue: $0$/$0.5$), GABA (orange: variable), Gln (red: $0.8$), Glu (green: $0.2$).}
  \label{fig:phantom_conc}
\end{figure}

The performance of the network is evaluated and compared with other state-of-the-art methods using a spectral benchmark dataset created by scanning calibrated phantoms of the metabolites studied here~\cite{Data}. The MEGA-PRESS spectra were acquired on a \SI{3}{\tesla} Siemens Magnetom Skyra system at Swansea University using the Siemens WIP MEGA-PRESS implementation with $T_E=\SI{68}{\milli\second}$ and $T_R=\SI{2000}{\milli\second}$ with $160$ averages per spectrum. Four datasets for phantoms of known composition, E1, E3 \& E4a and E4b, were used. E1, E3 and E4a were acquired with an acquisition bandwidth of \SI{1250}{\hertz}, while E4b was acquired at \SI{2000}{\hertz} (for the same phantoms as E4a). The published experimental dataset also contains a non-pH calibrated dataset E2, which was excluded as it was deemed not representative of the \invivo{} environment. E4c and E4d, also contained in the dataset, are repeat runs of E4a and E4b and we obtained similar results for them, so they are not further discussed here. Phantom composition is displayed in Table~\ref{table:phantom_conc} and Fig.~\ref{fig:phantom_conc}; for details of phantom preparation see~\cite{Jenkins2018, quantpaper}.

\section{Method}\label{sec:method}

We present the network structure and training procedure for MRSNet, our deep MRS quantification network, minimal data pre-processing and performance evaluation measures.

\subsection{Network Structure}

MRSNet is a 2D CNN. Input data to the network is a $N \times 2048$ matrix, where each row contains one component of a single spectrum. The maximum number of input rows is $N=9$ for the real and imaginary part and magnitude of up to three spectra (edit-off, edit-on and difference). Spectra are zero-padded and trimmed to obtain a length of $2048$ in the range from \SI{4.5}{ppm} to \SI{1.5}{ppm}, i.e. each column is a single frequency bin \SI{0.0014}{ppm} wide, providing ample resolution for small features. The majority of convolutions are $1 \times N$, spanning across the frequency bins to align with the separate rows of the input spectra. This network structure is loosely based on the CNN VGG architecture~\cite{Simonyan2014} and experimentally derived. Multiple layer 1D and 2D networks and single-layer 3D networks were explored by varying the arrangement of acquisitions and data type representations. However, they generally have poor performance compared to the suggested network structure.

The architecture of the network is designed to train identification of spectral features in individual rows (component of a single spectrum) early on, to be combined further down the network. These reduction layers are implemented in the middle of the network and repeated to combine the rows until the output tensor has one row. Using convolutions spanning multiple rows of the input early on lead to very poor results, whereas combining the convoluted rows near the middle of the network provided the best results.

\begin{table}
  \caption{Small network structure without pooling. Elipsis indicate repeated layers of the same class. Keys: conv: convolution; BN: batch normalisation; DO: drop out; FC: fully connected; ReLU: rectified linear unit.}\label{tab:network_arch}
  \begin{tabularx}{\columnwidth}{@{}lX@{}}
  \toprule
  \textbf{Layer class}    & \textbf{Description}\\
  \midrule
  input                   & $[1-9]\xtimes2048$ spectra\\
  conv1                   & $256\xtimes1\xtimes7$ conv, $1\xtimes2$ stride, ReLU, BN, DO$=0.4$\\
  conv2                   & $256\xtimes1\xtimes5$ conv, $1\xtimes2$ stride, ReLU, DO$=0.4$\\
  reduction1              & $256\xtimes[1-3]\xtimes[3]$ conv, ReLU, DO$=0.25$\\
  reduction1              & \dots\\
  \midrule
  conv3                   & $256\xtimes1\xtimes3$ conv, padding=same, ReLU, DO$=0.25$\\
  reduction2              & $256\xtimes1\xtimes3$ conv, $1\xtimes3$ stride, ReLU, DO$=0.25$\\
  conv3                   & \dots\\
  reduction2              & \dots\\
  \midrule
  conv4                   & $512\xtimes1\xtimes3$ conv, padding=same, ReLU, DO$=0.25$\\
  reduction3              & $512\xtimes1\xtimes3$ conv, 1x3 stride, ReLU, DO$=0.25$\\
  conv4                   & \dots\\
  reduction3              & \dots\\
  \midrule
  dense1                  & $1024$ FC Layer\\
  output                  & $5$ FC Layer, softmax\\
  \bottomrule
  \end{tabularx}
\end{table}

The complete network architecture is shown in Table~\ref{tab:network_arch} for the `small' variant without pooling. For every variant of the network, the layer class \emph{reduction1} is repeated until the output is a tensor with one row. For input spectra with nine rows, the layer is repeated five times, whereas for an input of a tensor with one row the layer has a convolutional kernel sized $3 \times 1$ and is repeated once. This is where the information contained in the different rows (spectra) of the input tensor are combined. A variety of substitutions are made to test different `medium' and `large' networks and pooling variants as shown in Tables~\ref{table:layer_sub} and~\ref{table:pooling_sub}.

The network is trained using the ADAM weight update~\cite{Kingma2014} with an experimentally derived learning rate of $10^{-4}$ and suggested beta values of $\beta_1=0.9$ and $\beta_2=0.999$ respectively. Mean squared error (MSE) is used as the loss function over mean absolute percentage error (MAPE) or mean absolute error (MAE) due to the higher overall performance seen experimentally. The training was performed on an NVIDIA Titan X, using Python 2.7.15, Keras 2.2.4~\cite{Keras} with Tensorflow 1.9.0 and CUDA 9.0.176.

\begin{table}
  \caption{Network layer class substitutions for `medium' and `large' networks, differences are in bold.}\label{table:layer_sub}
  \begin{tabularx}{\columnwidth}{@{}llX@{}}
  \toprule
  \textbf{Network}  & \textbf{Layer class}  & \textbf{Description}\\
  \midrule
  Medium            & conv1                 & $256\xtimes1\xtimes\mathbf{9}$ conv, $1\xtimes2$ stride,\\
                    &                       & ReLU, BN, DO$=0.4$\\
                    & conv2                 & $256\xtimes1\xtimes\mathbf{7}$ conv, $1\xtimes2$ stride,\\
                    &                       & ReLU, DO$=0.4$\\
                    & reduction1            & $256\xtimes[1-3]\xtimes\mathbf{5}$ conv, ReLU, DO$=0.25$ \\
  \midrule
  Large             & conv1                 & $256\xtimes1\xtimes\mathbf{16}$ conv, $1\xtimes2$ stride,\\
                    &                       & ReLU, BN, DO$=0.4$\\
                    & conv2                 & $256\xtimes1\xtimes\mathbf{8}$ conv, $1\xtimes2$ stride,\\
                    &                       & ReLU, DO$=0.4$\\
                    & reduction1            & $256\xtimes[1-3]\xtimes\mathbf{7}$ conv, ReLU, DO$=0.25$\\
  \bottomrule
  \end{tabularx}
\end{table}

\begin{table}
  \caption{Network substitutions for pooling variants where strided convoloutions are replaced by a MaxPool layer following a convoloution layer.}\label{table:pooling_sub}
  \begin{tabularx}{\columnwidth}{@{}llX@{}}
  \toprule
  \textbf{Network}    & \textbf{Layer class}  & \textbf{Description} \\
  \midrule
  All (pooling)       & reduction2            & $256\xtimes1\xtimes3$ conv, ReLU, DO$=0.25$\\
                      &                       & $1\xtimes3$ MaxPool \\
                      & reduction3            & $512\xtimes1\xtimes3$ conv, ReLU, DO$=0.25$\\
                      &                       & $1\xtimes3$ MaxPool \\
  \bottomrule
  \end{tabularx}
\end{table}

\begin{table*}
  \caption{Performance of network trained with LCModel basis set, three acquisitions, three data types using convolutions with strides.}\label{table:network_performance_strides}
  \begin{tabularx}{\textwidth}{@{}l|l|c|CC|*{4}{C}@{}}
  \toprule
  \textbf{Batch}  & \textbf{Network}  & \textbf{s/Epoch}  & \textbf{Train} & \textbf{Validation} & \textbf{E1} & \textbf{E3} & \textbf{E4a} & \textbf{E4b} \\
  \midrule
  16          & Small         & 51 &  0.00933 $\sigma$ 0.008 & 0.01113 $\sigma$ 0.011 & 0.12954 $\sigma$ 0.099 & 0.08898 $\sigma$ 0.075 & 0.07845 $\sigma$ 0.056 & 0.10344 $\sigma$ 0.080 \\
  16          & Medium        & 90 &  0.00858 $\sigma$ 0.008 & 0.01090 $\sigma$ 0.011 & 0.14440 $\sigma$ 0.126 & 0.08013 $\sigma$ 0.064 & 0.08157 $\sigma$ 0.055 & 0.11970 $\sigma$ 0.074 \\
  16          & Large         & 112 & \textbf{0.00769 $\sigma$ 0.007} & \textbf{0.00968 $\sigma$ 0.011} & 0.12680 $\sigma$ 0.106 & 0.07382 $\sigma$ 0.051 & 0.08439 $\sigma$ 0.057 & 0.12502 $\sigma$ 0.068 \\
  \midrule
  32          & Small         & 50 &  0.01096 $\sigma$ 0.010 & 0.01293 $\sigma$ 0.013 & 0.11980 $\sigma$ 0.099 & 0.08003 $\sigma$ 0.063 & 0.06619 $\sigma$ 0.053 & 0.13130 $\sigma$ 0.093 \\
  32          & Medium        & 84 &  0.00948 $\sigma$ 0.009 & 0.01194 $\sigma$ 0.012 & 0.14928 $\sigma$ 0.127 & 0.08752 $\sigma$ 0.070 & 0.09360 $\sigma$ 0.059 & 0.12649 $\sigma$ 0.076 \\
  32          & Large         & 100 & 0.01169 $\sigma$ 0.010 & 0.01409 $\sigma$ 0.013 & 0.17127 $\sigma$ 0.134 & \textbf{0.06996 $\sigma$ 0.043} & \textbf{0.06271 $\sigma$ 0.046} & 0.10637 $\sigma$ 0.081 \\
  \midrule
  \rowcolor{Gainsboro}
  64          & Small         & \textbf{48} &  0.01150 $\sigma$ 0.010 & 0.01274 $\sigma$ 0.012 & 0.11599 $\sigma$ 0.092 & 0.07378 $\sigma$ 0.060 & 0.07168 $\sigma$ 0.048 & 0.12210 $\sigma$ 0.090 \\
  64          & Medium        & 81 &  0.00961 $\sigma$ 0.008 & 0.01135 $\sigma$ 0.011 & 0.12914 $\sigma$ 0.111 & 0.09217 $\sigma$ 0.067 & 0.11139 $\sigma$ 0.049 & 0.13296 $\sigma$ 0.074 \\
  64          & Large         & 96 &  0.01348 $\sigma$ 0.011 & 0.01567 $\sigma$ 0.014 & \textbf{0.11031 $\sigma$ 0.097} & 0.11308 $\sigma$ 0.080 & 0.10906 $\sigma$ 0.093 & \textbf{0.08609 $\sigma$ 0.085} \\
  \bottomrule
  \end{tabularx}
\end{table*}

\begin{table*}
  \caption{Performance of network trained with LCModel basis set, three acquisitions, three data types using max pooling.}\label{table:network_performance_pooling}
  \begin{tabularx}{\textwidth}{@{}l|l|c|CC|*{4}{C}c@{}}
  \toprule
  \textbf{Batch}  & \textbf{Network}  & \textbf{s/Epoch}  &\textbf{Train} & \textbf{Validation} & \textbf{E1} & \textbf{E3} & \textbf{E4a} & \textbf{E4b} \\
  \midrule
  16          & Small         & 52 &  0.02319 $\sigma$ 0.020 & 0.02537 $\sigma$ 0.022 & 0.23780 $\sigma$ 0.183 & 0.12593 $\sigma$ 0.086 & 0.14338 $\sigma$ 0.100 & 0.16654 $\sigma$ 0.093 \\
  16          & Medium        & 91 &  \textbf{0.02155 $\sigma$ 0.018} & \textbf{0.02419 $\sigma$ 0.022} & 0.25608 $\sigma$ 0.203 & 0.12456 $\sigma$ 0.095 & 0.14460 $\sigma$ 0.097 & 0.18350 $\sigma$ 0.108 \\
  16          & Large         & 114 & 0.02955 $\sigma$ 0.024 & 0.03267 $\sigma$ 0.027 & 0.23548 $\sigma$ 0.203 & \textbf{0.10779 $\sigma$ 0.085} & 0.11319 $\sigma$ 0.088 & 0.15390 $\sigma$ 0.100 \\
  \midrule
  32          & Small         & 49 &  0.02550 $\sigma$ 0.021 & 0.02848 $\sigma$ 0.024 & 0.24714 $\sigma$ 0.194 & 0.14428 $\sigma$ 0.107 & 0.15845 $\sigma$ 0.115 & 0.16844 $\sigma$ 0.097 \\
  32          & Medium        & 85 &  0.02825 $\sigma$ 0.022 & 0.03044 $\sigma$ 0.025 & 0.26010 $\sigma$ 0.203 & 0.11713 $\sigma$ 0.084 & 0.11767 $\sigma$ 0.089 & 0.16157 $\sigma$ 0.086 \\
  32          & Large         & 101 & 0.02463 $\sigma$ 0.019 & 0.02652 $\sigma$ 0.022 & \textbf{0.23353 $\sigma$ 0.153} & 0.11028 $\sigma$ 0.061 & \textbf{0.10509 $\sigma$ 0.068} & \textbf{0.11382 $\sigma$ 0.074} \\
  \midrule
  \rowcolor{Gainsboro}
  64          & Small         & \textbf{48} &  0.02502 $\sigma$ 0.021 & 0.02701 $\sigma$ 0.024 & 0.25574 $\sigma$ 0.175 & 0.13671 $\sigma$ 0.103 & 0.14247 $\sigma$ 0.109 & 0.14566 $\sigma$ 0.089 \\
  64          & Medium        & 82 &  0.03168 $\sigma$ 0.025 & 0.03358 $\sigma$ 0.027 & 0.24458 $\sigma$ 0.197 & 0.12839 $\sigma$ 0.095 & 0.15139 $\sigma$ 0.105 & 0.14814 $\sigma$ 0.097 \\
  64          & Large         & 97 &  0.04081 $\sigma$ 0.028 & 0.04236 $\sigma$ 0.030 & 0.24200 $\sigma$ 0.196 & 0.11145 $\sigma$ 0.087 & 0.10594 $\sigma$ 0.082 & 0.15611 $\sigma$ 0.097 \\
  \bottomrule
  \end{tabularx}
\end{table*}

\subsection{Pre-processing}\label{sec:pre-processing}

All spectra are $B_0$ corrected w.r.t. the \SI{2}{ppm} NAA singlet peak due to NAA's low sensitivity to temperature of \SI{0.01}{ppm/\celsius}~\cite{Coman2009}. The experimental spectra are filtered using a first-order Butterworth filter to reduce noise and with no phase correction done for simplicity, as it can often require human interaction~\cite{DeBrouwer2009}. Experimental and simulated spectra are mean-centred and normalised so that the largest peak has amplitude $\pm 1$ across acquisitions with $x'=x/\max(|x|)$. Different acquisition bandwidths are dealt with by zero-filling the time domain signal to achieve a spectral resolution of $2048$ points in the selected \si{ppm} range.

\subsection{Performance Evaluation}

The error $\epsilon$ is calculated as the mean of the absolute differences of the actual $a_{i,j}$ and the predicted $p_{i,j}$ relative concentration for every label (metabolite) $j=1, \dotsc, L$, for every prediction (spectrum) $i=1, \dotsc, N$:
\begin{equation}\label{eq:performance_err}
  \epsilon = \frac{1}{NL}\sum^{N}_{i=1}\sum^{L}_{j=1} |a_{i,j}-p_{i,j}|.
\end{equation}
The standard deviation $\sigma$ is calculated in the usual way, $\sigma^2 = \sum^{N}_{i=1}\sum^{L}_{j=1}(p_{i,j}-\epsilon)^2 / (NL)$. This provides a good indication of overall network performance but is insensitive to low concentration metabolites. To counter this, an in-depth regression and MAPE analysis is performed for the best network on a per-metabolite basis in Section~\ref{sec:compare_with_current_methods}.

\section{Experiments}\label{sec:experiments}

We investigate a range of network structures, spectra, representations and basis sets to find a favourable combination. Network structure and batch sizes are investigated first, followed by the effect of different representations of the spectra and the choice of basis set. The LCModel basis set is used for the first two experiments as the analysis program has been shown to perform well by the comparative study in~\cite{Jenkins2018, quantpaper} and it performs best compared to other state-of-the-art quantification methods (see Table~\ref{table:performance_all_programs}), suggesting a good fit with experimental data. Training and validation are performed using simulated spectra as described in Section~\ref{sec:simulation}. Initial experiments in Sections~\ref{sec:network_structure} and~\ref{sec:data_type_and_channels} are trained for $100$ epochs, with $4,000$ training, $1,000$ validation samples. For the final experiment in Section~\ref{sec:data_source_comparison}, more compute time is dedicated to refining the networks over $200$ epochs with $5,000$ training samples. A light early stopping criteria is utilised throughout to prevent overfitting, with a minimum loss decrease of $10^{-12}$, a patience (number of epochs without improvement) of $15$ and `restore best weights' enabled. Network performance is benchmarked using experimental datasets from samples with known composition rather than performance on validation as described in Section~\ref{sec:benchmark_datasets}.

\subsection{Network Structure Investigation}\label{sec:network_structure}

An initial investigation is performed to explore the effect of minibatch sizes, convolutional kernel widths and pooling vs. stridden convolutions on performance with the benchmark datasets. Three variants of the network are tested (small, medium and large) over a range of minibatch sizes ($64$, $32$, $16$) for two major variants of each network, either using stridden convolutions or max-pooling. These two variants are explored as convolutions with strides have been shown to be advantageous over max-pooling methods~\cite{Springenberg2014}. Networks are trained using all three spectra (edit-off, edit-on and difference) and all three data representations (real, imaginary and magnitude/absolute value) stacked to form a $9\times2048$ input tensor for each sample. The results in Tables~\ref{table:network_performance_strides} and~\ref{table:network_performance_pooling} show the general trend that using convolutions with strides obtains a substantially higher performance over max-pooling. Networks with smaller convolutional kernels are substantially quicker than their larger counterparts as they are less computationally expensive to calculate. The `small' network, using stridden convolutions and a minibatch size of $64$ achieves a good balance between performance and training time on the benchmark datasets, making it the chosen network architecture for the following experiments.

\subsection{Effect of Different Representations of Spectra}\label{sec:data_type_and_channels}

Using MEGA-PRESS data provides a unique opportunity to explore how training networks on different combinations of acquired spectra for one scan (edit-off, edit-on and difference) affects performance. Additionally, three data types have been chosen to represent the spectrum by taking the real, imaginary or magnitude component of the frequency domain signal. These components are explored by using the same generated dataset for training and validation but considering different combinations of spectra and data types.

\begin{table*}
  \caption{Network performance for different combinations of acquired spectra (edit-off, edit-on, difference) and data types (real: R, imaginary: I, magnitude: M) using the `small' network with strided convoloutions and the LCModel basis for training and validation dataset generation.}\label{table:datatype_and_channels}
  \begin{tabularx}{\textwidth}{@{}ll|*{2}{C}|*{4}{C}X@{}}
  \toprule
  \textbf{Acquisitions}   & \textbf{Data type(s)}   &\textbf{Train} & \textbf{Validation}    & \textbf{E1}        & \textbf{E3}        & \textbf{E4a} & \textbf{E4b} \\
  \midrule
  Off                        & R                 &  0.01386 $\sigma$ 0.011 & 0.01521 $\sigma$ 0.014 & 0.13292 $\sigma$ 0.089 & 0.10046 $\sigma$ 0.080 & 0.08394 $\sigma$ 0.065 & 0.10424 $\sigma$ 0.084 \\
  Off                        & I                 &  0.01265 $\sigma$ 0.012 & 0.01494 $\sigma$ 0.014 & 0.13159 $\sigma$ 0.092 & 0.07989 $\sigma$ 0.054 & 0.09348 $\sigma$ 0.067 & 0.11512 $\sigma$ 0.070 \\
  Off                        & M                 &  0.01565 $\sigma$ 0.016 & 0.01949 $\sigma$ 0.024 & 0.06863 $\sigma$ 0.054 & 0.07800 $\sigma$ 0.063 & 0.06419 $\sigma$ 0.050 & 0.07662 $\sigma$ 0.054 \\
  Off                        & R I                 &  0.01203 $\sigma$ 0.011 & 0.01400 $\sigma$ 0.013 & 0.09873 $\sigma$ 0.064 & 0.10943 $\sigma$ 0.075 & 0.09235 $\sigma$ 0.067 & 0.08462 $\sigma$ 0.072 \\
  Off                        & R I M             &  \textbf{0.01105 $\sigma$ 0.010} & \textbf{0.01366 $\sigma$ 0.013} & 0.10024 $\sigma$ 0.062 & 0.09750 $\sigma$ 0.050 & 0.08600 $\sigma$ 0.059 & 0.09015 $\sigma$ 0.078 \\
  \midrule
  On                         & R                 &  0.01453 $\sigma$ 0.013 & 0.01695 $\sigma$ 0.017 & 0.22904 $\sigma$ 0.192 & 0.13183 $\sigma$ 0.124 & 0.11652 $\sigma$ 0.100 & 0.17211 $\sigma$ 0.077 \\
  On                         & I                 &  0.01256 $\sigma$ 0.012 & 0.01570 $\sigma$ 0.017 & 0.22129 $\sigma$ 0.191 & 0.13696 $\sigma$ 0.133 & 0.12930 $\sigma$ 0.119 & 0.19410 $\sigma$ 0.112 \\
  On                         & M                 &  0.01944 $\sigma$ 0.019 & 0.02567 $\sigma$ 0.030 & 0.21743 $\sigma$ 0.200 & 0.15283 $\sigma$ 0.100 & 0.15292 $\sigma$ 0.098 & 0.16059 $\sigma$ 0.095 \\
  On                         & R I               &  0.01369 $\sigma$ 0.012 & 0.01648 $\sigma$ 0.018 & 0.23515 $\sigma$ 0.182 & 0.13795 $\sigma$ 0.130 & 0.13242 $\sigma$ 0.114 & 0.19135 $\sigma$ 0.101 \\
  On                         & R I M             &  0.01468 $\sigma$ 0.014 & 0.01851 $\sigma$ 0.019 & 0.23231 $\sigma$ 0.194 & 0.17241 $\sigma$ 0.135 & 0.18429 $\sigma$ 0.101 & 0.19757 $\sigma$ 0.095 \\
  \midrule
  Diff                       & R                 &  0.04300 $\sigma$ 0.041 & 0.04888 $\sigma$ 0.047 & 0.08484 $\sigma$ 0.073 & 0.09208 $\sigma$ 0.059 & 0.09231 $\sigma$ 0.071 & 0.10738 $\sigma$ 0.081 \\
  Diff                       & I                 &  0.03993 $\sigma$ 0.040 & 0.04649 $\sigma$ 0.046 & 0.06341 $\sigma$ 0.066 & 0.11511 $\sigma$ 0.066 & 0.11253 $\sigma$ 0.070 & 0.10329 $\sigma$ 0.092 \\
  Diff                       & M                 &  0.04142 $\sigma$ 0.041 & 0.05125 $\sigma$ 0.050 & 0.04967 $\sigma$ 0.063 & 0.08806 $\sigma$ 0.069 & 0.08591 $\sigma$ 0.066 & 0.10017 $\sigma$ 0.074 \\
  Diff                       & R I               &  0.03963 $\sigma$ 0.040 & 0.04671 $\sigma$ 0.047 & 0.07428 $\sigma$ 0.066 & 0.07755 $\sigma$ 0.047 & 0.07148 $\sigma$ 0.052 & 0.08196 $\sigma$ 0.063 \\
  Diff                       & R I M             &  0.03918 $\sigma$ 0.040 & 0.04647 $\sigma$ 0.047 & 0.07753 $\sigma$ 0.079 & 0.06767 $\sigma$ 0.046 & 0.08112 $\sigma$ 0.057 & 0.09246 $\sigma$ 0.061 \\
  \midrule
  Off On                     & R                 &  0.01396 $\sigma$ 0.011 & 0.01626 $\sigma$ 0.014 & 0.20498 $\sigma$ 0.159 & 0.18993 $\sigma$ 0.146 & 0.16938 $\sigma$ 0.135 & 0.14621 $\sigma$ 0.082 \\
  Off On                     & I                 &  0.01425 $\sigma$ 0.012 & 0.01640 $\sigma$ 0.014 & 0.23126 $\sigma$ 0.170 & 0.16219 $\sigma$ 0.128 & 0.15234 $\sigma$ 0.120 & 0.19090 $\sigma$ 0.095 \\
  Off On                     & M                 &  0.01415 $\sigma$ 0.012 & 0.01642 $\sigma$ 0.016 & 0.14484 $\sigma$ 0.118 & 0.11244 $\sigma$ 0.086 & 0.12549 $\sigma$ 0.078 & 0.13234 $\sigma$ 0.070 \\
  Off On                     & R I               &  0.01434 $\sigma$ 0.013 & 0.01636 $\sigma$ 0.015 & 0.18232 $\sigma$ 0.135 & 0.16390 $\sigma$ 0.093 & 0.15403 $\sigma$ 0.095 & 0.15630 $\sigma$ 0.072 \\
  Off On                     & R I M             &  0.01618 $\sigma$ 0.013 & 0.01801 $\sigma$ 0.016 & 0.15784 $\sigma$ 0.132 & 0.17001 $\sigma$ 0.130 & 0.12872 $\sigma$ 0.105 & 0.12797 $\sigma$ 0.080 \\
  \midrule
  Off Diff                   & R                 &  0.01361 $\sigma$ 0.011 & 0.01440 $\sigma$ 0.013 & 0.06866 $\sigma$ 0.053 & 0.08917 $\sigma$ 0.051 & 0.07371 $\sigma$ 0.051 & 0.09721 $\sigma$ 0.076 \\
  Off Diff                   & I                 &  0.01203 $\sigma$ 0.011 & 0.01358 $\sigma$ 0.013 & 0.07983 $\sigma$ 0.053 & 0.04921 $\sigma$ 0.028 & 0.05918 $\sigma$ 0.042 & 0.10686 $\sigma$ 0.081 \\
  \rowcolor{Gainsboro}
  Off Diff                   & M                 &  0.01549 $\sigma$ 0.013 & 0.01759 $\sigma$ 0.017 & \textbf{0.05956 $\sigma$ 0.042} & \textbf{0.04519 $\sigma$ 0.031} & \textbf{0.04288 $\sigma$ 0.034} & \textbf{0.06034 $\sigma$ 0.054} \\
  Off Diff                   & R I               &  0.01292 $\sigma$ 0.011 & 0.01461 $\sigma$ 0.013 & 0.11167 $\sigma$ 0.090 & 0.05455 $\sigma$ 0.043 & 0.04936 $\sigma$ 0.037 & 0.10697 $\sigma$ 0.082 \\
  Off Diff                   & R I M             &  0.01344 $\sigma$ 0.012 & 0.01509 $\sigma$ 0.015 & 0.07955 $\sigma$ 0.057 & 0.11050 $\sigma$ 0.074 & 0.09290 $\sigma$ 0.059 & 0.10666 $\sigma$ 0.085 \\
  \midrule
  On Diff                    & R                 &  0.01704 $\sigma$ 0.014 & 0.01819 $\sigma$ 0.016 & 0.20824 $\sigma$ 0.177 & 0.14069 $\sigma$ 0.106 & 0.14328 $\sigma$ 0.103 & 0.17063 $\sigma$ 0.080 \\
  On Diff                    & I                 &  0.01439 $\sigma$ 0.012 & 0.01552 $\sigma$ 0.014 & 0.19468 $\sigma$ 0.161 & 0.13248 $\sigma$ 0.087 & 0.13502 $\sigma$ 0.084 & 0.17786 $\sigma$ 0.098 \\
  On Diff                    & M                 &  0.01888 $\sigma$ 0.018 & 0.02156 $\sigma$ 0.022 & 0.15265 $\sigma$ 0.131 & 0.10383 $\sigma$ 0.078 & 0.12364 $\sigma$ 0.076 & 0.12842 $\sigma$ 0.056 \\
  On Diff                    & R I               &  0.01673 $\sigma$ 0.014 & 0.01840 $\sigma$ 0.016 & 0.22594 $\sigma$ 0.191 & 0.15966 $\sigma$ 0.118 & 0.15320 $\sigma$ 0.106 & 0.19460 $\sigma$ 0.086 \\
  On Diff                    & R I M             &  0.01957 $\sigma$ 0.019 & 0.02295 $\sigma$ 0.022 & 0.23033 $\sigma$ 0.184 & 0.16489 $\sigma$ 0.111 & 0.17125 $\sigma$ 0.101 & 0.17671 $\sigma$ 0.091 \\
  \midrule
  Off On Diff                & R                 &  0.01326 $\sigma$ 0.011 & 0.01486 $\sigma$ 0.013 & 0.13909 $\sigma$ 0.086 & 0.09442 $\sigma$ 0.067 & 0.08806 $\sigma$ 0.058 & 0.10422 $\sigma$ 0.083 \\
  Off On Diff                & I                 &  0.01251 $\sigma$ 0.010 & 0.01393 $\sigma$ 0.012 & 0.16848 $\sigma$ 0.112 & 0.08873 $\sigma$ 0.081 & 0.09467 $\sigma$ 0.081 & 0.14555 $\sigma$ 0.085 \\
  Off On Diff                & M                 &  0.01336 $\sigma$ 0.011 & 0.01517 $\sigma$ 0.014 & 0.07679 $\sigma$ 0.055 & 0.06491 $\sigma$ 0.053 & 0.06009 $\sigma$ 0.040 & 0.06971 $\sigma$ 0.046 \\
  Off On Diff                & R I               &  0.01624 $\sigma$ 0.013 & 0.01710 $\sigma$ 0.015 & 0.15457 $\sigma$ 0.107 & 0.13011 $\sigma$ 0.067 & 0.14073 $\sigma$ 0.074 & 0.15130 $\sigma$ 0.086 \\
  Off On Diff                & R I M             &  0.01418 $\sigma$ 0.012 & 0.01612 $\sigma$ 0.014 & 0.14309 $\sigma$ 0.104 & 0.08350 $\sigma$ 0.073 & 0.09529 $\sigma$ 0.070 & 0.11978 $\sigma$ 0.081 \\
  \bottomrule
  \end{tabularx}
\end{table*}

Results in Table~\ref{table:datatype_and_channels} show a clear performance advantage of utilising the magnitude representation for all acquisition types. This is expected, as there is a level of phase uncertainty from the experimental signal leading to a potential disagreement with the real and phase spectra from the benchmark set when compared to the basis set. However, using the magnitude spectra does increase the linewidth when compared to only the real spectrum; which is typically preferred for quantification for this reason. Despite this, the networks prefer the more predictable magnitude spectra, at the cost of dealing with the increased linewidth.

The best combination of acquisitions is the edit-off and difference, as these provide the maximal amount of information with no repetition of data. The reduced performance for the difference only acquisition can be attributed to the Cr spectra, wherein the LCModel basis set that was used to train the network assumes perfect editing, where the normal and inverted spectrum in the edit-on and edit-off acquisition match amplitude. As such, there is no residual Cr signal in the difference spectrum (which is not always the case in practice), as it has been edited out leading to difficulties in quantification. Networks perform worse when redundant data is supplied, for example, in the case of using the edit-off, on and difference magnitude spectra. This could be because the network is provided with too many degrees of freedom, increasing the training difficulty. For the following experiments, we choose networks that use the magnitude edit-off and difference spectra as inputs.

\subsection{Choice of Basis Set}\label{sec:data_source_comparison}

We compare the performance of three basis sets from LCModel, PyGamma and FID-A as described in Section~\ref{sec:simulation} as differences in the basis sets may have a significant impact on quantification performance~\cite{Jenkins2018, quantpaper}. Networks are trained and tested using the magnitude of the edit-off and difference spectrum with an increased number of training samples $5,000$, $1,000$ validation and test spectra, and with an increased number of $200$ epochs for training.

Datasets were generated using the same metabolite concentration and noise values across the three basis sets to create a like-for-like comparison. Results in Table~\ref{table:basis_set_compare} suggest that for overall quantification, the single linewidth (\SI{1}{\hertz}) FID-A basis performs the best for E1, E4a and E4b, while the LCModel basis outperforms the others for the E3 dataset. The PyGamma basis performs substantially worse overall. This suggests that overall the FID-A basis set is the best choice for training the network for general quantification.

\begin{table*}
  \caption{A comparison of basis set influence on MRSNet performance, using the `small' network with stridden convoloutions. Networks are trained using the magnitude of the edit-off and difference acqusition. MLW denotes that the dataset is comprised of multiple linewidth spectra.}\label{table:basis_set_compare}
  \begin{tabularx}{\textwidth}{@{}l|CC|CCCC@{}}
  \toprule
  \textbf{Basis} & \textbf{Train} & \textbf{Validation} & \textbf{E1} & \textbf{E3} & \textbf{E4a} & \textbf{E4b} \\
  \midrule
  LCModel & 0.01206 $\sigma$ 0.009 & 0.01547 $\sigma$ 0.017 & 0.07936 $\sigma$ 0.061 & \textbf{0.04553 $\sigma$ 0.035} & 0.04991 $\sigma$ 0.038 & 0.06702 $\sigma$ 0.046 \\
  \rowcolor{Gainsboro}
  FID-A & 0.01120 $\sigma$ 0.009 & 0.01374 $\sigma$ 0.013 & \textbf{0.04744 $\sigma$ 0.035} & 0.05323 $\sigma$ 0.036 & \textbf{0.04670 $\sigma$ 0.033} & \textbf{0.05960 $\sigma$ 0.049} \\
  FID-A MLW & \textbf{0.00995 $\sigma$ 0.008} & \textbf{0.01308 $\sigma$ 0.013} & 0.04773 $\sigma$ 0.035 & 0.05963 $\sigma$ 0.041 & 0.05407 $\sigma$ 0.040 & 0.06302 $\sigma$ 0.051 \\
  PyGamma & 0.01337 $\sigma$ 0.012 & 0.01688 $\sigma$ 0.017 & 0.09081 $\sigma$ 0.071 & 0.09496 $\sigma$ 0.057 & 0.09396 $\sigma$ 0.064 & 0.10269 $\sigma$ 0.066 \\
  PyGamma MLW & 0.01133 $\sigma$ 0.009 & 0.03436 $\sigma$ 0.037 & 0.08465 $\sigma$ 0.061 & 0.07521 $\sigma$ 0.058 & 0.06994 $\sigma$ 0.062 & 0.08147 $\sigma$ 0.064 \\
  \bottomrule
  \end{tabularx}
\end{table*}

We also investigate the effect of training the networks on basis sets generated with multiple linewidths (MLW). In practice, the linewidth of the experimental spectra is variable, depending on a multitude of environmental factors as described in Section~\ref{sec:mrs background}. Spectra are simulated using basis sets with linewidths of \SI{0.75}{\hertz}, \SI{1}{\hertz} and \SI{1.25}{\hertz} for the PyGamma and FID-A basis sets. For LCModel this was not possible as there is only one fixed linewidth basis set and we do not have access to the underlying generator. The network is trained using the same values for metabolite concentrations and noise as the single linewidth experiment. The generated spectra are split evenly between the defined linewidths. Table~\ref{table:basis_set_compare} shows that the overall accuracy of the networks trained with PyGamma improves but is marginally worse for FID-A. The use of simulators over fixed basis sets could be advantageous here as training a network on a range of linewidths should allow it to generalise to broader linewidths, typically seen in practice. However, in this instance, it appears that the linewidth of \SI{1}{\hertz} for FID-A closely matches the experimental spectra and using multiple linewidths has provided no improvement.

\section{Evaluation}\label{sec:evaluation}

Overall the best performing network is the `small' variant with stridden convolutions, using the magnitude of the edit-off and difference spectra. We evaluate its overall performance and compare it with state-of-the-art methods.

\subsection{Comparison with State-of-the-art Methods}\label{sec:compare_with_current_methods}

Unlike previous sections where performance was measured against all five metabolites, the performance in this section is evaluated on reduced sets of metabolites (see Table~\ref{table:performance_all_programs}). The sets are based on which metabolites are present in the phantom and an intersection of metabolites the selected programs can report. Cr is omitted from the analysis as it is not reported by LCModel for MEGA-PRESS difference analysis. Glu and Gln values are combined and reported as GLX as they are generally considered unresolvable at \SI{3}{\tesla} due to their similar chemical structure and resulting spectrum. For the reduced metabolite sets, reported concentration values are re-scaled to $c'_r = c_r / \sum_{r\in R} c_r$ where $R$ is the index set of the reduced metabolite set.

Quantification software settings, results and procedures are available in~\cite{Jenkins2018, quantpaper}. The error measures have been changed to focus on individual metabolite error for this article rather than the NAA/GABA ratio.

\begin{table*}
  \caption{Performance of state-of-the-art quantification programs compared to the best network (`small' with stridden convolutions). Multiple networks have been trained with different basis sets, including multiple linewidth bases (MLW), with results evaluated on reduced metabolite sets to enable a like-for-like comparison with the external analysis programs.}\label{table:performance_all_programs}
  \protected\def\B#1{\bfseries#1}
  \begin{threeparttable}
  \begin{tabularx}{\textwidth}{@{}l*{4}{C}@{}}
  \toprule
                & \textbf{E1} & \textbf{E3}           & \textbf{E4a}       & \textbf{E4b}       \\
  \midrule
  \textbf{Metabolite set} & [NAA, GABA] & [NAA, GABA, GLX] & [NAA, GABA, GLX] & [NAA, GABA, GLX] \\
  \midrule
  \textbf{Analysis program}\\
  \hspace{.8em} VeSPA   & 0.1192 $\sigma$ 0.096 & 0.1199 $\sigma$ 0.081 & 0.1552 $\sigma$ 0.105 & 0.2016 $\sigma$ 0.122 \\
  \hspace{.8em} TARQUIN & 0.1125 $\sigma$ 0.093 & 0.0708 $\sigma$ 0.049 & 0.0705 $\sigma$ 0.070 & 0.1341 $\sigma$ 0.090 \\
  \hspace{.8em} LCModel & 0.0462 $\sigma$ 0.030 & 0.0956 $\sigma$ 0.041 & 0.0919 $\sigma$ 0.054 & 0.1266 $\sigma$ 0.095 \\
  \hspace{.8em} jMRUI (AQUES)$^{\dagger}$  & 0.2081 $\sigma$ 0.092 & 0.1432 $\sigma$ 0.088 & 0.1865 $\sigma$ 0.117 & - \\
  \hspace{.8em} jMRUI (QUEST) $^{\dagger}$ & 0.2044 $\sigma$ 0.098 & 0.1402 $\sigma$ 0.080 & 0.1958 $\sigma$ 0.144 & - \\
  \hspace{.8em} Gannet  & 0.1405 $\sigma$ 0.103 & 0.3092 $\sigma$ 0.158 & 0.3012 $\sigma$ 0.179 & 0.5334 $\sigma$ 0.603 \\
  \midrule
  \textbf{MRSNet} trained with\\
  \rowcolor{Gainsboro}
  \hspace{.8em}LCModel    & \B{0.0318 $\sigma$ 0.014} & \B{0.0321 $\sigma$ 0.021 } & \B{0.0402 $\sigma$ 0.022} & 0.0720 $\sigma$ 0.050\\
  \hspace{.8em}FID-A      & 0.0746 $\sigma$ 0.044 & 0.0617 $\sigma$ 0.034 & 0.0402 $\sigma$ 0.033 & 0.0735 $\sigma$ 0.060 \\
  \hspace{.8em}FID-A MLW  & 0.0636 $\sigma$ 0.032 & 0.0714 $\sigma$ 0.044 & 0.0722 $\sigma$ 0.041 & 0.0864 $\sigma$ 0.054 \\
  \hspace{.8em}PyGamma    & 0.0408 $\sigma$ 0.021 & 0.0842 $\sigma$ 0.057 & 0.0674 $\sigma$ 0.041 & 0.0994 $\sigma$ 0.063 \\
  \hspace{.8em}PyGamma MLW & 0.0501 $\sigma$ 0.022 & 0.0708 $\sigma$ 0.044 & 0.0679 $\sigma$ 0.044 & \B{0.0704 $\sigma$ 0.055} \\
  \bottomrule
  \end{tabularx}
  \begin{tablenotes}
  \small
  \item $\dagger$ Quantification was not performed for the E4b dataset.
  \item LCModel, VeSPA and jMRUI report individual Glu and Gln values which have been combined into GLX. Gannet and TARQUIN report the combined GLX.
  \end{tablenotes}
  \end{threeparttable}
\end{table*}

The results in Table~\ref{table:performance_all_programs} show that overall MRSNet is more accurate and precise than other methods for quantification using MEGA-PRESS. In contrast to the results in Table~\ref{table:basis_set_compare}, training the network with the LCModel basis set provides the best performance in this instance. This performance increase is due to the reduced set of metabolites for analysis. However, when evaluating MRSNet performance in the general case across all metabolites the FID-A basis set outperforms the LCModel basis on average.

\begin{figure*}
  \includegraphics[width=.25\linewidth]{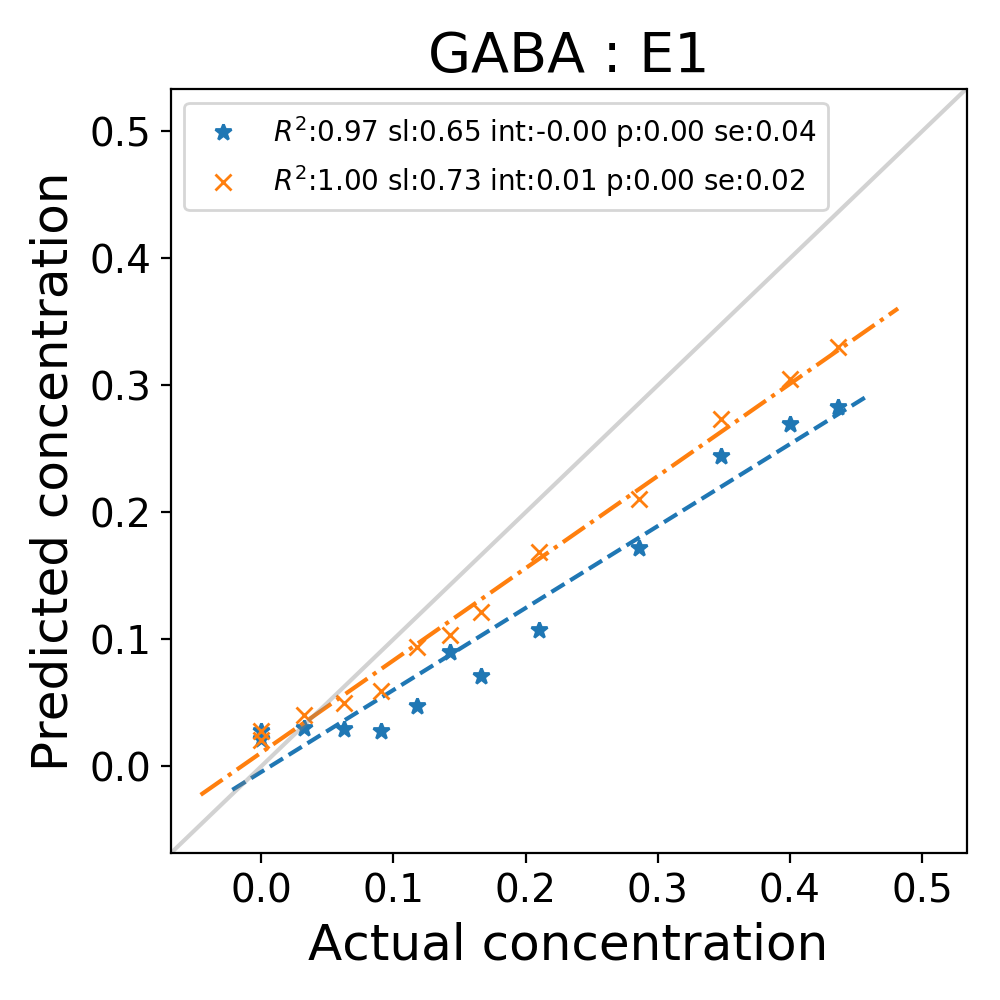}\hfill
  \includegraphics[width=.25\linewidth]{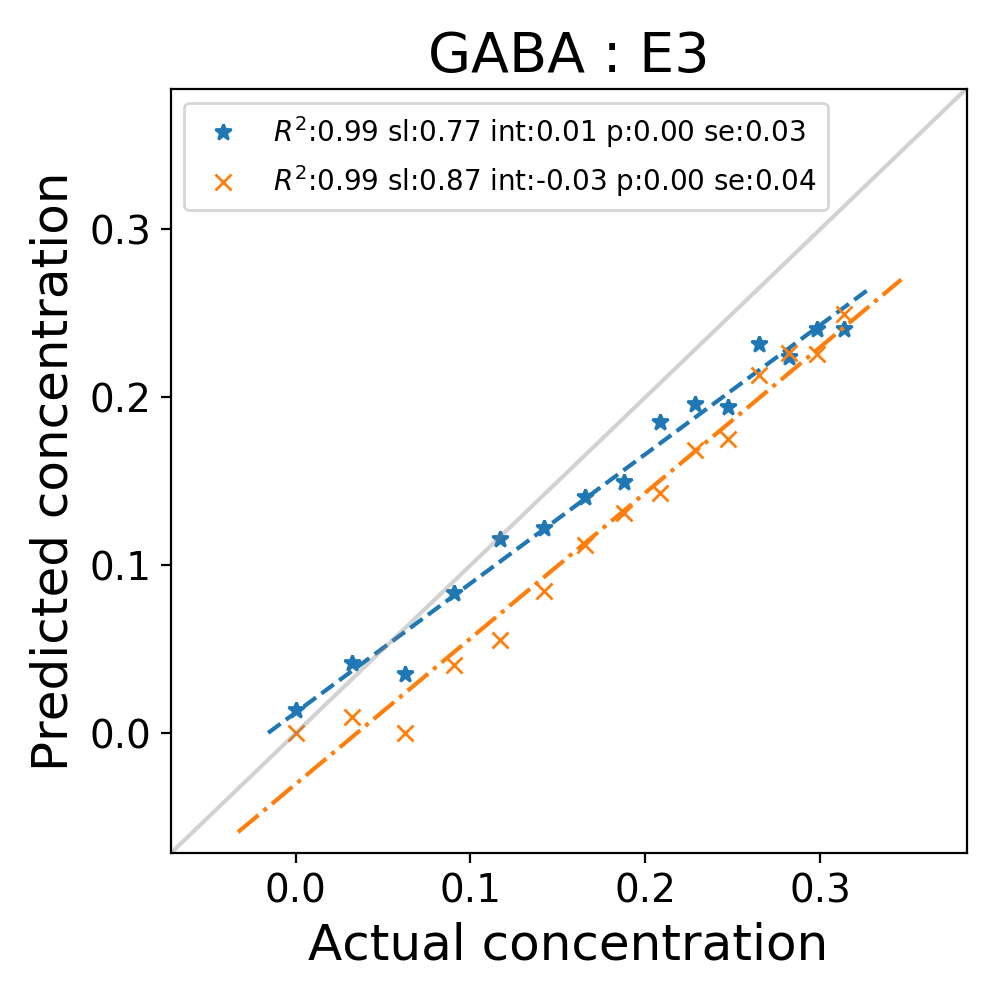}\hfill
  \includegraphics[width=.25\linewidth]{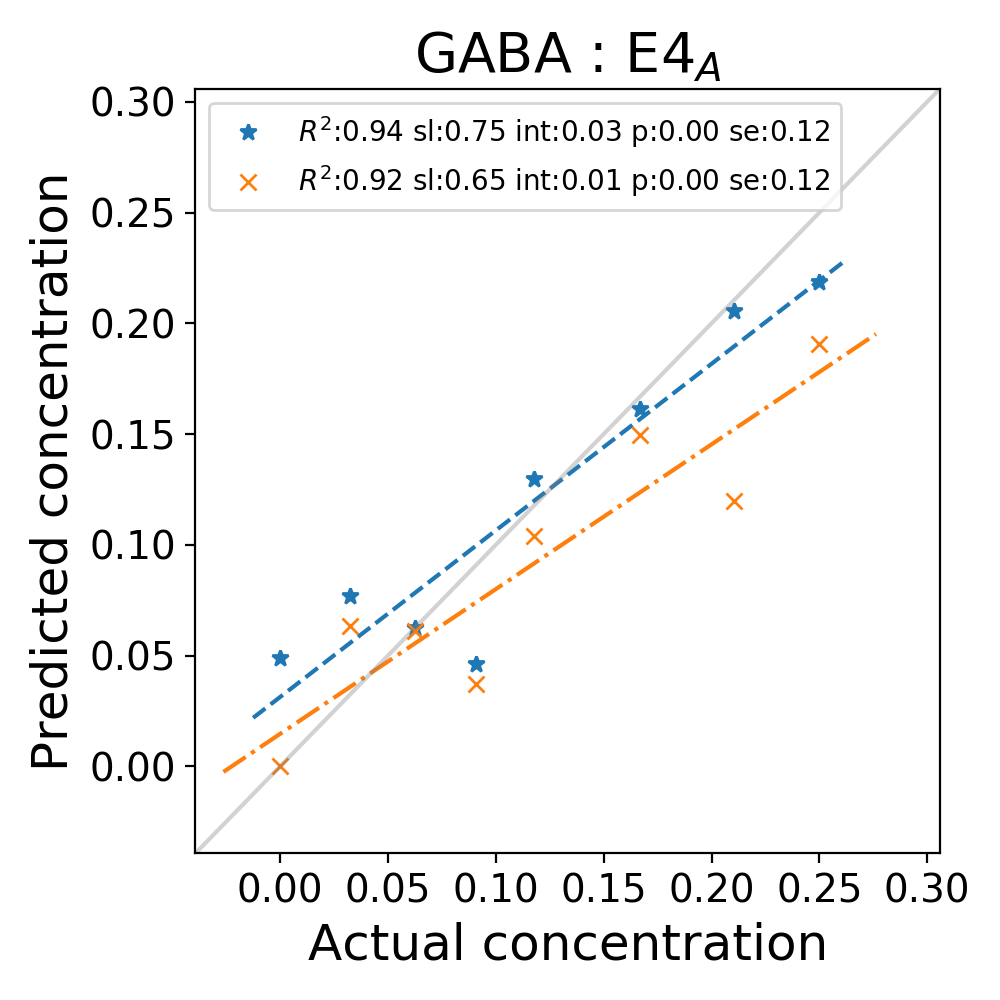}\hfill
  \includegraphics[width=.25\linewidth]{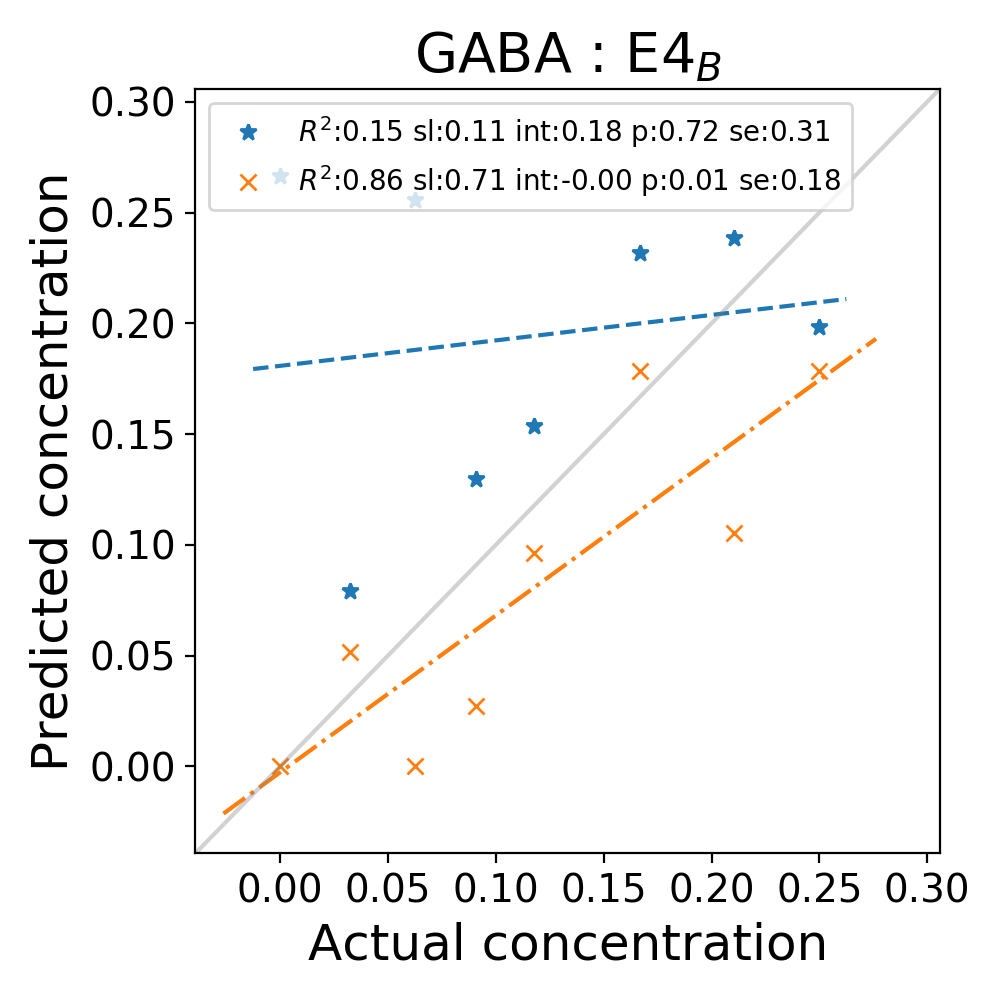}\hfill
  \par\medskip
  \includegraphics[width=.25\linewidth]{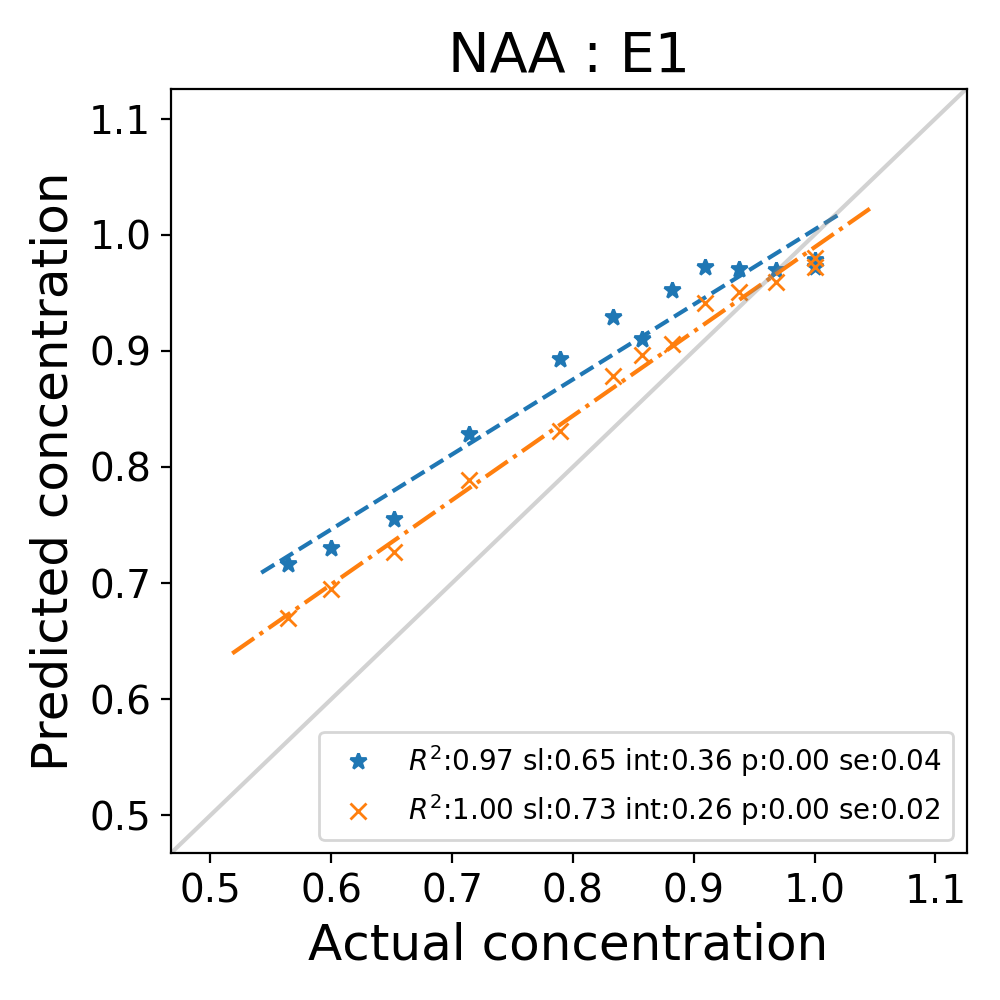}\hfil
  \includegraphics[width=.25\linewidth]{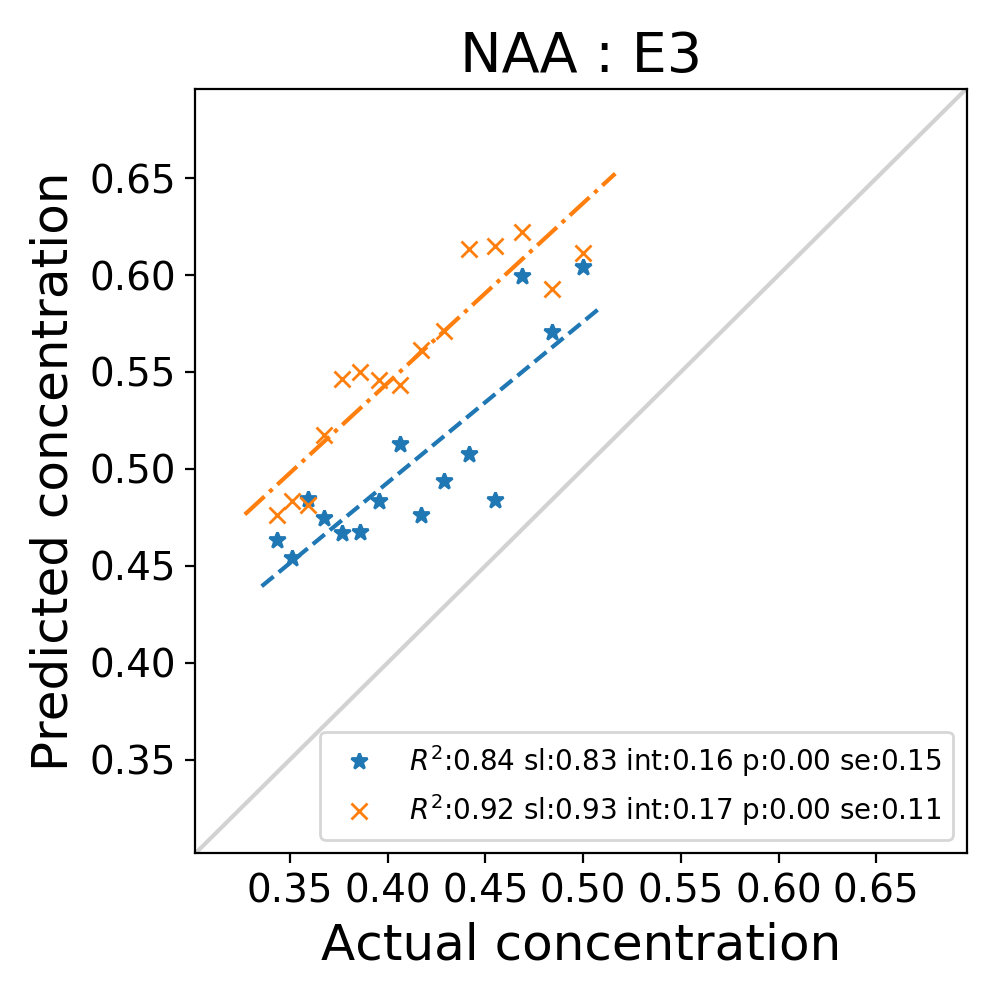}\hfil
  \includegraphics[width=.25\linewidth]{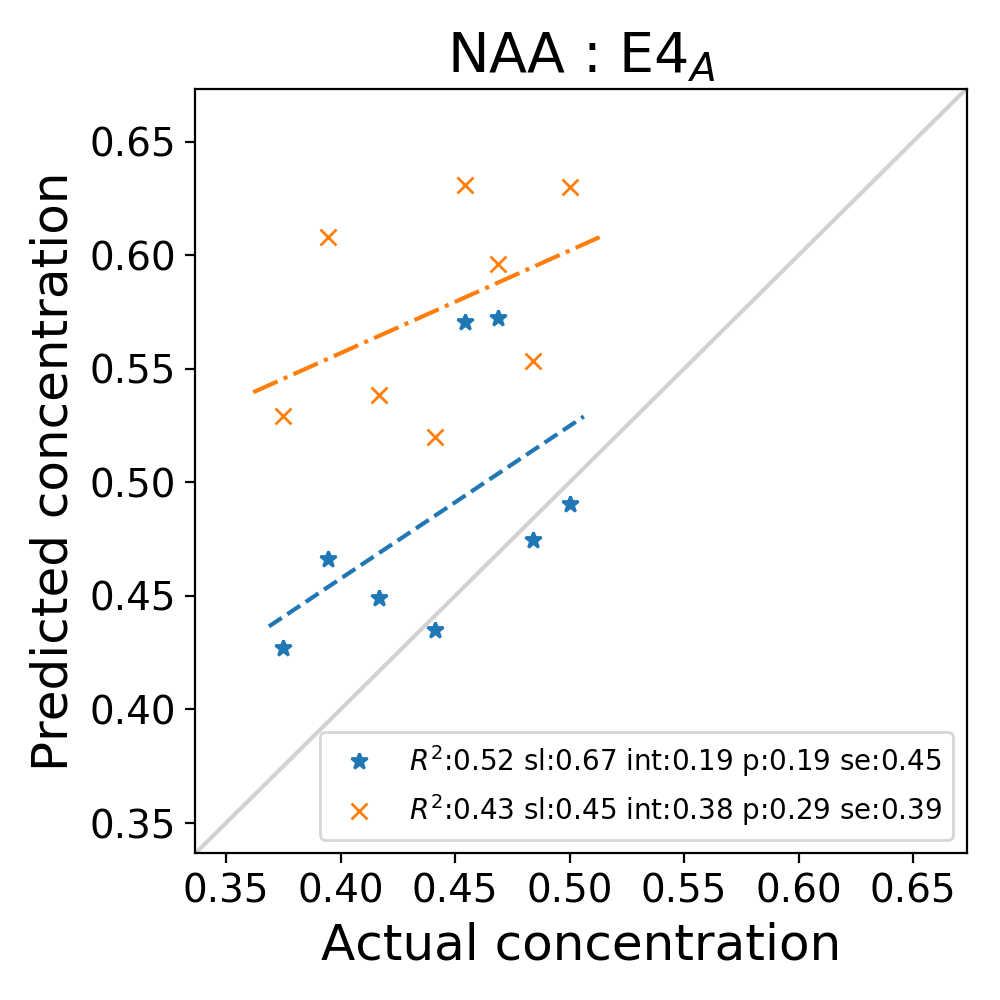}\hfil
  \includegraphics[width=.25\linewidth]{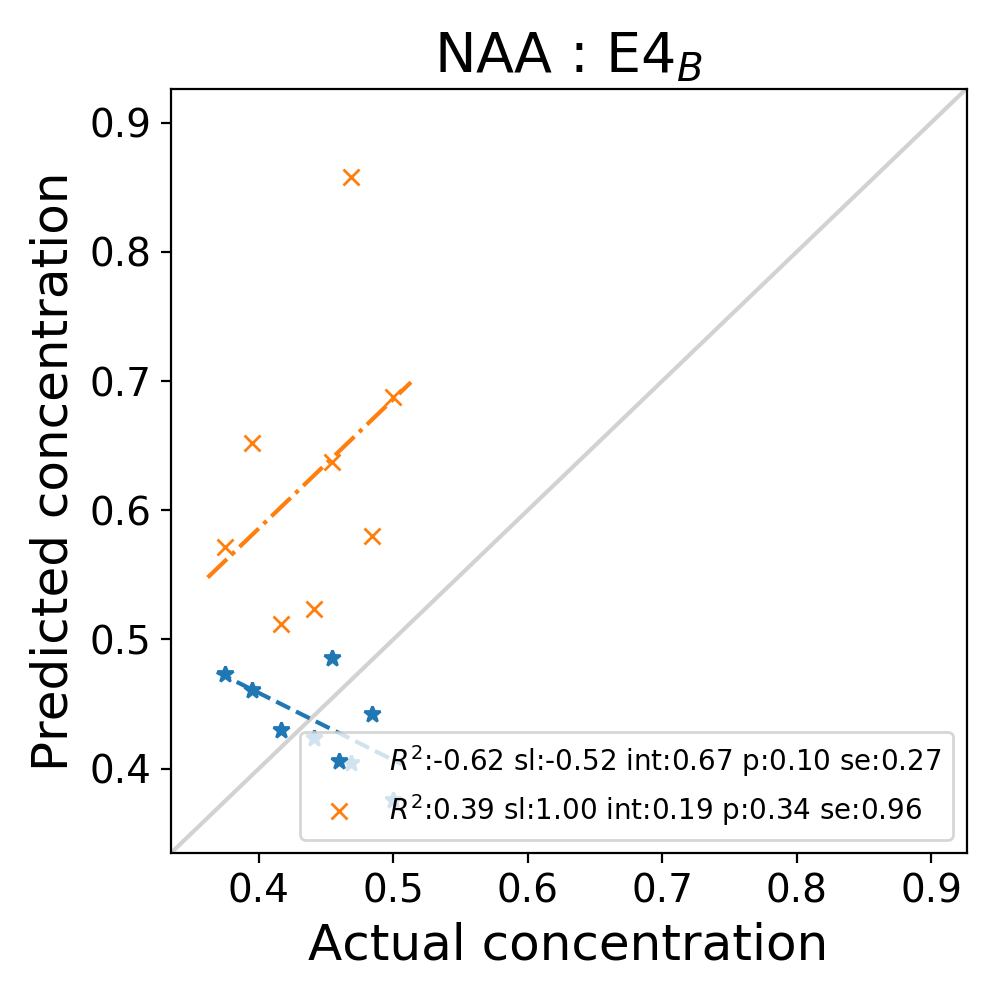}\hfil
  \par\medskip
  \caption{Regression analysis for the best network architecture (`small', stridden convoloutions) trained using the LCModel basis set (blue,~$\star$) compared with LCModel analysis (orange,~$\times$) for GABA (top row) and NAA (bottom row) across all benchmark datasets (from left to right, E1, E3, E4a, E4b). R$^2$, Slope (sl), intercept (int), p-value (p), and standard error (se) values are displayed on each graph in the legend. The ideal quantification method would have a slope of $1$ and an intercept at $0$, as represented by the faint grey diagonal line.}\label{fig:regression_analysis}
\end{figure*}

To further understand and analyse the performance of the network, regression analysis is performed for GABA and NAA across all benchmark datasets for LCModel and MRSNet trained with the LCModel basis due the performance advantage shown in Table~\ref{table:performance_all_programs}. Results in Fig.~\ref{fig:regression_analysis} show that the performance of MRSNet is comparable to LCModel, except for E4b, where both struggle. The E4b dataset is considerably noisier than the others and the performance may be due to LCModel's pre-processing steps. Analysis of all series shows that only E4b suffers from inversion of the edit-on or off spectrum, resulting in a poor difference spectrum, caused by the MRI scanner software (also see~\cite{quantpaper}).

To further investigate, individual metabolite error and MAPE values are calculated in Table~\ref{table:mape_metabolite_error}. A similar trend continues, where MRSNet outperforms LCModel on average, except for GABA quantification which bounces between LCModel and MRSNet trained with the FID-A `MLW' basis, suggesting that the GABA model from FID-A may be a better fit to experimental data.

The large error seen in Table~\ref{table:performance_all_programs} can be attributed to the performance of NAA quantification, as it has the highest concentration in all benchmark phantoms. Any improvement seen in NAA quantification has the largest impact on the error in Eq.~\eqref{eq:performance_err} and is reflected in the overall performance in Table~\ref{table:performance_all_programs}. While NAA is not typically the target of edited MRS, it can be utilised as an internal reference compound. So any improvement in the accuracy of NAA would indirectly improve quantification for all other metabolites.

\begin{table*}
  \caption{Individual metabolite MAPE for state-of-the-art quantification programs compared to the best network (`small' with stridden convolutions). Multiple networks have been trained with different basis sets, including multiple linewidth bases (MLW). See Table~\ref{table:performance_all_programs} for omission explanations.}\label{table:mape_metabolite_error}
  \protected\def\B#1{\bfseries#1}
  \addtolength{\tabcolsep}{-3.96pt}
  \begin{tabularx}{\textwidth}{@{}l|CC|CCC|CCC|CCC@{}}
  \toprule
   & \multicolumn{2}{c|}{\textbf{E1}} & \multicolumn{3}{c|}{\textbf{E3}} & \multicolumn{3}{c|}{\textbf{E4a}}  & \multicolumn{3}{c}{\textbf{E4b}}\\
  \midrule
   & NAA & GABA & NAA & GABA & GLX & NAA & GABA & GLX & NAA & GABA & GLX \\
  \midrule
  \textbf{Program}           & & & & & & & & & & \\
  \hspace{.8em}TARQUIN       & 16.26\% & 69.98\% & 10.64\% & 47.63\% & 18.62\% & 13.04\% & 158.19\% & 28.55\% & 7.93\% & 100.93\% & 27.17\%\\
  \hspace{.8em}VeSPA         & 16.96\% & 66.64\% & 43.52\% & 49.96\% & 23.00\% & 48.17\% & 82.69\% & 43.79\% & 54.42\% & 217.30\% & 55.48\%\\
  \hspace{.8em}LCModel       & 6.53\% & \B{24.62\%} & 35.35\% & 34.68\% & 21.82\% & 31.08\% & 35.20\% & 24.61\% & 42.49\% & \B{39.08\%} & 33.62\%\\
  \hspace{.8em}jMRUI (AQSES) & 24.12\% & 221.87\% & 18.65\% & 133.39\% & 40.34\% & 18.37\% & 312.78\% & 57.14\% 	& \multicolumn{3}{c}{-}\\
  \hspace{.8em}jMRUI (QUEST) & 23.47\% & 215.59\% & 12.90\% & 134.90\% & 47.68\% & 21.50\% & 332.47\% & 57.12\% 	& \multicolumn{3}{c}{-}\\
  \hspace{.8em}Gannet        & 20.13\% & 76.41\% & 114.12\% & 57.56\% & 87.59\% & 102.48\% & 44.97\% & 90.06\%	& 99.92\% & 139.99\% & 135.18\%\\
  \midrule
  \textbf{MRSNet} with       & & & & & & & & & & \\
  \rowcolor{Gainsboro}
  \hspace{.8em}LCModel       & \B{3.89\%} & 25.29\% & \B{4.59\%} & 27.93\% & \B{10.53\%} & \B{5.15\%} & 52.59\% & 13.30\% & 11.22\% & 104.94\% & 17.08\%\\
  \hspace{.8em}FID-A         & 10.37\% & 42.56\% & 22.63\% & 18.20\% & 15.02\% & 11.60\% & 30.97\% & \B{10.42\%} & 13.00\% & 85.83\% & \B{16.41\%}\\
  \hspace{.8em}FID-A MLW     & 8.71\% & 36.90\% & 25.87\% & \B{11.45\%} & 20.08\% & 22.85\% & \B{29.17\%} & 19.70\% & 20.05\% & 84.95\% & 21.46\%\\
  \hspace{.8em}PyGamma       & 5.18\% & 30.33\% & 14.13\% & 57.37\% & 29.63\% & 5.71\% & 123.52\% & 19.85\% & 13.35\% & 160.74\% & 22.14\%\\
  \hspace{.8em}PyGamma MLW 	& 6.14\% & 44.89\% & 13.15\% & 53.56\% & 23.65\% & 8.61\% & 83.27\% & 22.63\% & \B{7.00\%} & 87.71\% & 22.52\%\\
  \bottomrule
  \end{tabularx}
\end{table*}

\subsection{Basis Comparison}\label{sec:basis_comparison}

The variety of performance shown in the basis comparison experiment and subsequent MAPE analysis (see Tables~\ref{table:basis_set_compare} and~\ref{table:mape_metabolite_error}) can be explained by the variance of spectra generated from three data sources. As mentioned in Section~\ref{sec:simulation}, dataset and simulator creators have a large range of options available to them for simulation parameters, along with a range of metabolite models, resulting in a large range of possible spectra.

\begin{figure}
  \centering
  \includegraphics[width=0.95\columnwidth]{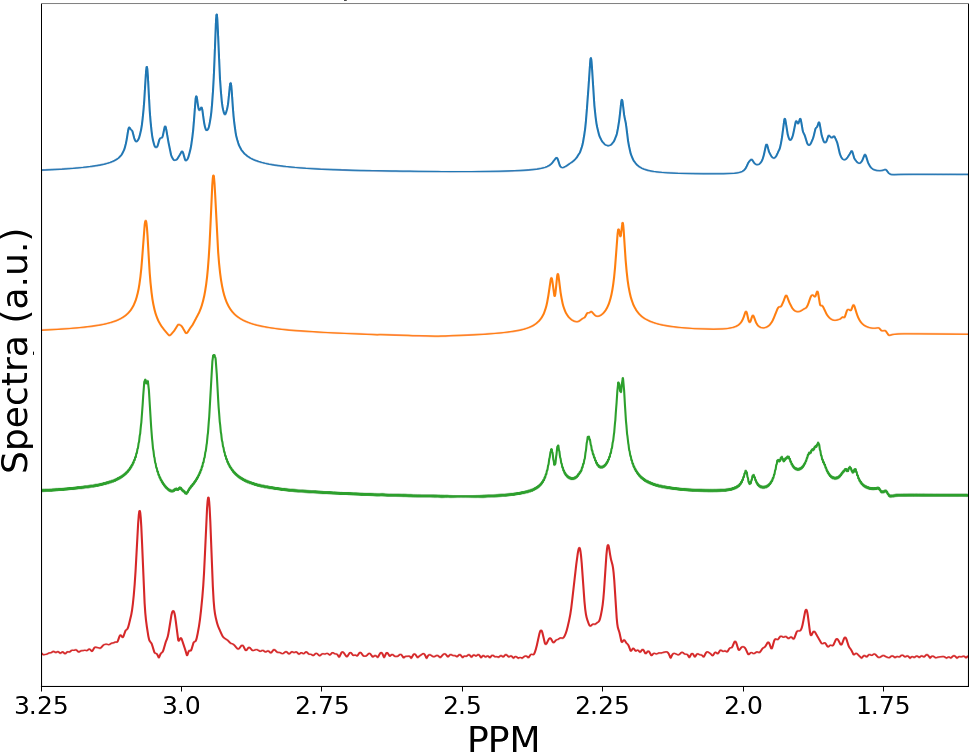}
  \caption{Example of spectra differences for the magnitude MEGA-PRESS difference GABA spectra from multiple sources, from top to bottom: PyGamma, FID-A, LCModel, experimental spectrum.}\label{fig:gaba_compare}
\end{figure}

Fig.~\ref{fig:gaba_compare} compares the difference spectra from PyGamma, FID-A and LCModel with an additional in-house experimentally acquired pure GABA phantom spectrum using MEGA-PRESS. From visual inspection, it is clear to see that none of the simulated data sources aligns perfectly with the experimental spectrum. This may be due to a multitude of reasons, including each simulator using a different GABA model: PyGamma uses Govindaraju \emph{et al.}~\cite{Govindaraju2000}, FID-A uses Near \emph{et al.}~\cite{Urban2010} and LCModel uses Kaiser \emph{et al.}~\cite{Song2008}. All simulators use Govindaraju \emph{et al.}'s~\cite{Govindaraju2000} values for the other metabolite models. A different GABA model is often chosen due to the well-known issue that the values in~\cite{Govindaraju2000} are rough approximations of the true values~\cite{Kreis2012}. In addition to simulation parameters, there is a multitude of experimental factors that can alter the resulting spectrum, further complicating the issue of matching simulation to experimental results. In an ideal scenario, networks should be trained on a large range of experimental data to cover the potential variation of spectra. However, in practice, this is a non-trivial task, with phantom creation being difficult and time-consuming in addition to needing a large amount of data to cover the potential experimental variations of the spectra.

\subsection{Benefits and Limitations}

MRSNet is quick to train on GPUs (18 minutes on an NVIDIA Titan X GPU), with a relatively low number of samples ($5,000$). Once trained, it requires no special hardware and it takes on average \SI{24}{\pm2\milli\second} to process a single spectrum with the network, using a dual-core \SI{2.7}{\giga\hertz} i7-7500u CPU, with the network occupying 95MB disk space. Furthermore, it requires no interaction from a user with specialist knowledge for potentially complicated processing steps, such as phase-correction.

The data type and channel experiments in Section~\ref{sec:data_type_and_channels} is not generally applicable to all basis sets in Table~\ref{table:basis_set_compare}, as this study was only performed using the LCModel basis set. As shown, other basis sets have different spectra and may have a more accurate representation of real or phase data, but it is expected that the magnitude spectra will remain the best performing due to the issue of uncertainty in phase reconstruction from the scanner.

Networks in this paper are only trained and tested with one timing variant of a single pulse sequence (MEGA-PRESS at \SI{3}{\tesla} with $T_R = \SI{68}{\milli\second}$, $T_E = \SI{2000}{\milli\second}$) with a benchmark dataset collected from one scanner, and for a single frequency window (\SI{4.5}{ppm} to \SI{1.5}{ppm}). This network should generalise to different MEGA-PRESS spectra, but it is unlikely that it will work as accurately with different strength $B_0$ fields, scanner manufacturers, pulse sequences or pulse sequence timings without further training.

Finally, \invivo{} data has not been used to evaluate this method due to the lack of ground truth data, which would, of course, be very difficult to obtain. Additionally, we have chosen to explore a limited number of metabolic signals, while there is a much larger range of spectra and macromolecule signals that are obtainable with MRS \invivo{}.

\section{Conclusions and Future Work}

We have demonstrated that the overall accuracy and precision of metabolite quantification in MRS is improved by a convolutional neural network by comparing its performance to current state-of-the-art methods. We have found that a 2D CNN, using stridden convolutions, and utilising `small' $1\xtimes N$ convoloutions along the frequency axis of the input spectra is the best performing network architecture.

Additionally, we have explored a range of data sources for training and representations for the input spectra. This has shown that the edit-off and difference absolute spectra provide the best results for the benchmark datasets. Furthermore, in the general case, training the network should be done with the FID-A basis, however, when looking at a reduced set of metabolites that excludes Creatine, the LCModel basis outperforms the alternatives. This highlights the non-trivial task of basis set selection and the need for more accurate characterisation.

There is a range of directions for future work, as this study focuses on the frequency domain for a single pulse sequence at one $T_E$, $T_R$ timing for five metabolites. Different representations of the input data could be investigated, such as using the complex time-domain signal, multiple $T_E$ acquisitions, multiple short-time Fourier transforms or reduced frequency domain data such as peak locations, amplitudes and phases. Networks could be trained and tested for a larger number of metabolites and a range of pulse sequences and timings to see how well they can learn to generalise. Importantly, such work must continue to be linked to experimental work with calibrated phantoms to try to ensure accuracy and reliability \invivo{}.

\ifCLASSOPTIONcompsoc
  \section*{Acknowledgments}
\else
  \section*{Acknowledgment}
\fi

The authors would like to thank the Experimental MRI Center (EMRIC, Cardiff University) for the use of their LCModel license. MC would like to thank EPSRC and Cardiff University for doctoral funding.

\ifCLASSOPTIONcaptionsoff
  \newpage
\fi

\bibliographystyle{IEEEtran-abbrev}
\bibliography{library}

\end{document}